\documentclass[sigconf]{acmart}
\usepackage{seqsplit}
\usepackage{amsmath,amsfonts}
\usepackage{graphicx}
\usepackage{textcomp}
\usepackage{xcolor}
\usepackage[many]{tcolorbox}
\usepackage{booktabs}
\usepackage{multirow}
\usepackage{array}

\newcolumntype{C}[1]{>{\centering\arraybackslash}p{#1}}

\usepackage{newtxtext}
\usepackage{newtxmath}

\usepackage{algorithm}
\usepackage{algpseudocode}

\usepackage{hyperref}

\usepackage{listings}

\lstdefinelanguage{XML}
{
  morestring=[b]",
  morestring=[s]{>}{<},
  morecomment=[s]{<!--}{-->},
  identifierstyle=\color{blue},
  keywordstyle=\color{cyan},
  morekeywords={xmlns,version,type}
}

\usepackage{footnote}
\usepackage{xspace}
\newcommand{\system}{{\sc AdvDroidZero}\xspace}
\newcommand{\paragraphbe}[1]{\smallskip\noindent{\bf {#1}.}~}

\usepackage[abbreviations]{foreign}

\AtBeginDocument{%
  \providecommand\BibTeX{{%
    \normalfont B\kern-0.5em{\scshape i\kern-0.25em b}\kern-0.8em\TeX}}}

\copyrightyear{2023}
\acmYear{2023}
\setcopyright{acmlicensed}
\acmConference[CCS '23] {Proceedings of the 2023 ACM SIGSAC Conference on Computer and Communications Security}{November 26--30, 2023}{Copenhagen, Denmark.}
\acmBooktitle{Proceedings of the 2023 ACM SIGSAC Conference on Computer and Communications Security (CCS '23), November 26--30, 2023, Copenhagen, Denmark}
\acmPrice{15.00}
\acmISBN{979-8-4007-0050-7/23/11}
\acmDOI{10.1145/3576915.3623117}

\definecolor{grey}{HTML}{d3d3d3}
\newtcolorbox{boxA}{
    boxrule = 1.5pt,
    rounded corners,
    colback = grey, 
    boxsep=0.1pt,
}

\usepackage{etoolbox}
\makeatletter
\patchcmd{\authornote}{\g@addto@macro\addresses{\@authornotemark}}{}{}{}
\makeatother

\begin{document}

\title{Efficient Query-Based Attack against ML-Based Android Malware Detection under Zero Knowledge Setting}


\author{Ping He}
\affiliation{%
  \institution{Zhejiang University}
  \country{}
  }
\email{gnip@zju.edu.cn}

\author{Yifan Xia}
\affiliation{%
  \institution{Zhejiang University}
  \country{}
  }
\email{yfxia@zju.edu.cn}

\author{Xuhong Zhang}
\affiliation{%
  \institution{Zhejiang University}
  \country{}
  }
\email{zhangxuhong@zju.edu.cn}

\author{Shouling Ji}
\authornote{Corresponding Author}
\authornotemark[1]
\affiliation{%
  \institution{Zhejiang University}
  \country{}
  }
\email{sji@zju.edu.cn}


\begin{abstract}
The widespread adoption of the Android operating system has made malicious Android applications an appealing target for attackers.
Machine learning-based (ML-based) Android malware detection (AMD) methods are crucial in addressing this problem; however, their vulnerability to adversarial examples raises concerns. 
Current attacks against ML-based AMD methods demonstrate remarkable performance but rely on strong assumptions that may not be realistic in real-world scenarios, e.g., the knowledge requirements about feature space, model parameters, and training dataset.
To address this limitation, we introduce \system, an efficient query-based attack framework against ML-based AMD methods that operates under the zero knowledge setting.
Our extensive evaluation shows that \system is effective against various mainstream ML-based AMD methods, in particular, state-of-the-art such methods and real-world antivirus solutions.
\end{abstract}



\maketitle

\section{Introduction}

The Android operating system is prevalent in today's technologically advanced world, powering various mobile devices and applications~\cite{MobileMarket}.
However, its popularity has also led to a significant issue: Android malware.
Recent security reports have identified over 33 million Android malware samples, posing a considerable threat to users' privacy and data integrity \cite{MalwareSample,DBLP:conf/uss/SunSLM21,DBLP:conf/ndss/ShenVS21,DBLP:conf/uss/ShenVS22,DBLP:journals/tdsc/Suarez-TangilS22,DBLP:conf/sp/ChatterjeeDOHPF18}.
To combat this, researchers have widely adopted machine learning-based (ML-based) Android malware detection (AMD) methods \cite{DBLP:conf/ndss/ArpSHGR14,DBLP:conf/ccs/ZhangZZDCZZY20,DBLP:conf/ndss/MaricontiOACRS17,DBLP:journals/tissec/DaoudiABK22,DBLP:journals/csur/LiuTLL23,DBLP:journals/comsur/FarukiBLGGCR15}.
These techniques efficiently identify and classify potential threats by analyzing Android application features, making ML-based AMD effective in protecting users and their devices.

Unfortunately, ML-based AMD techniques are vulnerable to attacks \cite{DBLP:journals/corr/SzegedyZSBEGF13,DBLP:conf/sp/Carlini017,DBLP:conf/sp/ChenJW20,DBLP:conf/iclr/BrendelRB18,DBLP:conf/uss/Suya20,DBLP:conf/ndss/VoAR22,DBLP:conf/uss/FuD00022}.
To evade them, adversaries can manipulate malicious applications in the problem space, causing the feature vector to cross the ML classifier's decision boundary.
With such an attack, adversaries can effectively generate thousands of realistic, inconspicuous adversarial Android malware at scale, making ``adversarial malware as a service'' a real threat.
Although various attacks \cite{DBLP:conf/esorics/GrossePMBM17,DBLP:conf/acsac/YangK0G17,DBLP:journals/tifs/LiL20,DBLP:conf/sp/PierazziPCC20,DBLP:journals/tifs/0002LWW0N0020,DBLP:conf/ccs/ZhaoZZZZLYYL21} against ML-based AMD methods have been proposed, they assume that the adversary has extensive knowledge of the target system, including access to the dataset, feature space, and model parameters.
For instance, the HRAT attack \cite{DBLP:conf/ccs/ZhaoZZZZLYYL21} understands the feature space of the target system and consequently designs four types of function call manipulation to modify the function call graph.
Besides, it additionally utilizes gradient information to choose the manipulation type and the manipulated function, necessitating knowledge of the dataset and model parameters.
This assumption might be too strong and not always applicable in real-world situations.
In practice, adversaries often have limited or no knowledge about the target system, making the system's internals, such as the dataset, feature space, and model parameters, unknown.

The efficient generation of adversarial Android malware under zero knowledge setting remains an open question.
However, addressing this problem is far from straightforward, primarily due to two critical challenges: the \textbf{vast heterogeneous perturbation space} originating from the sample side and the \textbf{zero knowledge challenge} stemming from the detection side.

The first challenge is the numerous heterogeneous perturbations possible for the Android application package (APK) file, attributed to its complex structure.
The APK file is a ZIP archive, typically containing a manifest file and Dalvik Executable (DEX) codes.
The adversary's modification space for APK files is more diverse than that for images or text.
For example, within an APK, an adversary could modify permissions in the manifest file or alter function call relations in DEX codes, while in images, they would only need to change pixel values.
In addition, even when considering homogeneous perturbations, the perturbation space is still large and discrete.
For instance, if the adversary wants to add a permission element in the manifest file, there are over 150 permissions with different protection levels in the Android system, each with varying evasion effectiveness.

Another challenge in this scenario is that the adversary has zero knowledge about the target ML-based AMD method.
Different target methods may use different heterogeneous features, e.g., permission, activities, function call graphs, etc.
As a result, carefully designed perturbations (e.g., modifying function call relations only) may not align with the target system's features (e.g., permissions only), making them ineffective.
For instance, changing the function call relations is useless for Drebin~\cite{DBLP:conf/ndss/ArpSHGR14} since Drebin does not use the function call-related feature and thus the elaborate perturbations in the problem space do not affect it at all.
Consequently, employing transfer attacks \cite{DBLP:conf/ccs/PapernotMGJCS17,DBLP:conf/sp/MaoFWJ0LZLB022,DBLP:conf/icpads/ZhangZLWDG21} directly may prove to be suboptimal due to the feature space gap.
Even if the adversary attempts to deduce the features of the target model, they can only rely on query feedback, i.e., labels and confidence, which is inadequate for efficiently identifying the target system's precise features.
Furthermore, previous works~\cite{DBLP:journals/tifs/LiL20, DBLP:journals/tifs/0002LWW0N0020,DBLP:conf/ccs/ZhaoZZZZLYYL21} expect to acquire real-time fine-grained feedback from the target model, e.g., the gradients of the target model, to direct efficient and effective attack.
Moreover, under the zero knowledge setting, the adversary can only get coarse-grained feedback, i.e., labels or confidence, which makes the attack process blind.
Therefore, such an attack process is not applicable because the adversary only has zero knowledge in this scenario.

To address the aforementioned challenges, we propose a query-based \textbf{adv}ersarial An\textbf{droid} malware attack framework under \textbf{zero} knowledge setting, termed \system.
At a high level, \system proposes a novel data structure termed perturbation selection tree to mitigate the challenge of vast heterogeneous perturbation space and a semantic-based adjustment policy to tackle the zero knowledge challenge.
Our intuition is that perturbations in the APK file have varying levels of evasion effectiveness and exhibit semantic meaning.
Perturbations with similar semantics are more likely to display comparable evasion effectiveness.
For instance, adding a \textit{uses-feature} element with \textit{android.hardware.audio.output} value in the manifest file means the perturbation is related to hardware and audio.
The perturbation will have similar evasion effectiveness with adding a \textit{uses-feature} element with \textit{android.hardware.audio.pro} value.
Once a perturbation with positive evasion effectiveness is identified, we can infer other perturbations with positive evasion effectiveness based on semantic meaning, thereby facilitating the selection of optimal perturbations. 

Based on these insights, \system designs a perturbation selection tree based on semantics at varying granularities, abstracting the perturbation selection process into path sampling in the tree.
Similar to the decision tree models in machine learning that can handle heterogeneous data types, the perturbation selection tree can naturally manage heterogeneous perturbations.
To address the vast perturbation space challenge, \system clusters semantically similar perturbations and executes them simultaneously.
This approach reduces the perturbation search space, enhancing the efficiency of the generation process.
To tackle the zero knowledge challenge, \system designs a semantic-based adjustment policy to change a batch of node probabilities based on their abstract semantic levels in cascade.
The high-level semantics representing the perturbation type (e.g., perturbations in manifest or DEX codes) can be determined through the cascade process by a few initial queries. 
Meanwhile, the semantic-based adjustment policy can diffuse the query feedback information from the selected perturbation to its semantically similar perturbations through the tree structure.
Hence, this policy can adjust semantically similar perturbations simultaneously via every single query.
As a result, the semantic-based adjustment policy efficiently increases the probabilities of clusters with positive evasion effectiveness while reducing those with negative evasion effectiveness under the zero knowledge setting.

We evaluate \system using the large Android malware dataset containing 135,859 benign applications and 15,778 malware (151,637 total).
This evaluation demonstrates the attack performance of \system against various mainstream ML-based AMD methods using static program analysis approaches, including Drebin \cite{DBLP:conf/ndss/ArpSHGR14}, Drebin-DL \cite{DBLP:conf/esorics/GrossePMBM17}, APIGraph \cite{DBLP:conf/ccs/ZhangZZDCZZY20} and MaMadroid \cite{DBLP:conf/ndss/MaricontiOACRS17}.
\system outperforms the baseline methods \cite{DBLP:conf/asiaccs/SongLAGKY22} in terms of both attack effectiveness and runtime overheads.
In particular, \system achieves about 90\% attack success rate against these mainstream ML-based AMD methods.
Additionally, \system requires only around 15 queries to generate most of (80\%) adversarial Android malware and completes the process within 10 minutes.
Furthermore, \system also demonstrates effectiveness against real-world antivirus solutions (AVs), sandboxes in VirusTotal, and the state-of-the-art (SOTA) robust AMD method FD-VAE \cite{DBLP:conf/www/LiZYLGC21}.
Specifically, \system reduces the number of detected antivirus engines for approximately 92\% of malware samples within 10 queries against VirusTotal and decreases the number of detected antivirus engines by 47.98\% per successfully attacked malware. 
Moreover, AdvDroidZero also achieves about 89\% attack success rate within 10 queries against the VirusTotal on the recent malware samples from 2022.
Even when targeting the sandboxes in VirusTotal and the state-of-the-art robust AMD methods, \system still maintains the attack success rate of 37\% under 10 query budgets and 77\% under 30 query budgets, respectively.

The key contributions of this paper are summarized as follows.
We propose \system, which is an efficient query-based attack framework for generating adversarial Android malware under zero knowledge setting.
In \system, a novel data structure called perturbation selection tree is proposed to address the vast heterogeneous search space challenge, and a semantic-based adjustment policy is proposed to choose the corresponding perturbations efficiently.

\section{Background}

\subsection{Android Application Package}

APK is the file format used for distributing and installing applications on the Android operating system.
The APK file structure aims to encompass all the necessary aspects of an application, from its core code and resources to its metadata.
It typically consists of several key components: \textit{AndroidManifest.xml}, \textit{classes.dex}, \textit{resources.arsc}, \textit{res/}, \textit{assets/}, \textit{META-INF/}, \textit{lib/}.
Among these, \textit{AndroidManifest.xml} and \textit{classes.dex} play pivotal roles within the APK.
The \textit{AndroidManifest.xml} file carries comprehensive configuration information about the APK, while the \textit{classes.dex} file holds the program semantics of the application.
Additional details regarding these components can be found in Appendix~\ref{appendix:apk}.

The mainstream ML-based AMD methods \cite{DBLP:conf/ndss/ArpSHGR14,DBLP:conf/ndss/MaricontiOACRS17,DBLP:conf/ccs/ZhangZZDCZZY20} typically extract features from the manifest file and DEX codes (i.e., \textit{AndroidManifest.xml} and \textit{classes.dex}).
As an informative configuration file, \textit{AndroidManifest.xml} specifies Android application requirements, e.g., permission to access the Internet.
The syntax features (static features) stemming from the manifest file are usually represented as binary features, where the value indicates whether a certain element is declared or not.
\textit{classes.dex} assembles the source codes of an Android application.
The semantic features (dynamic features) extracted from \textit{classes.dex} contain rich program semantics, e.g., the function call graph captures the function relation semantics.

\begin{figure}[t]   
	\centering  
	\includegraphics[width=1\linewidth]{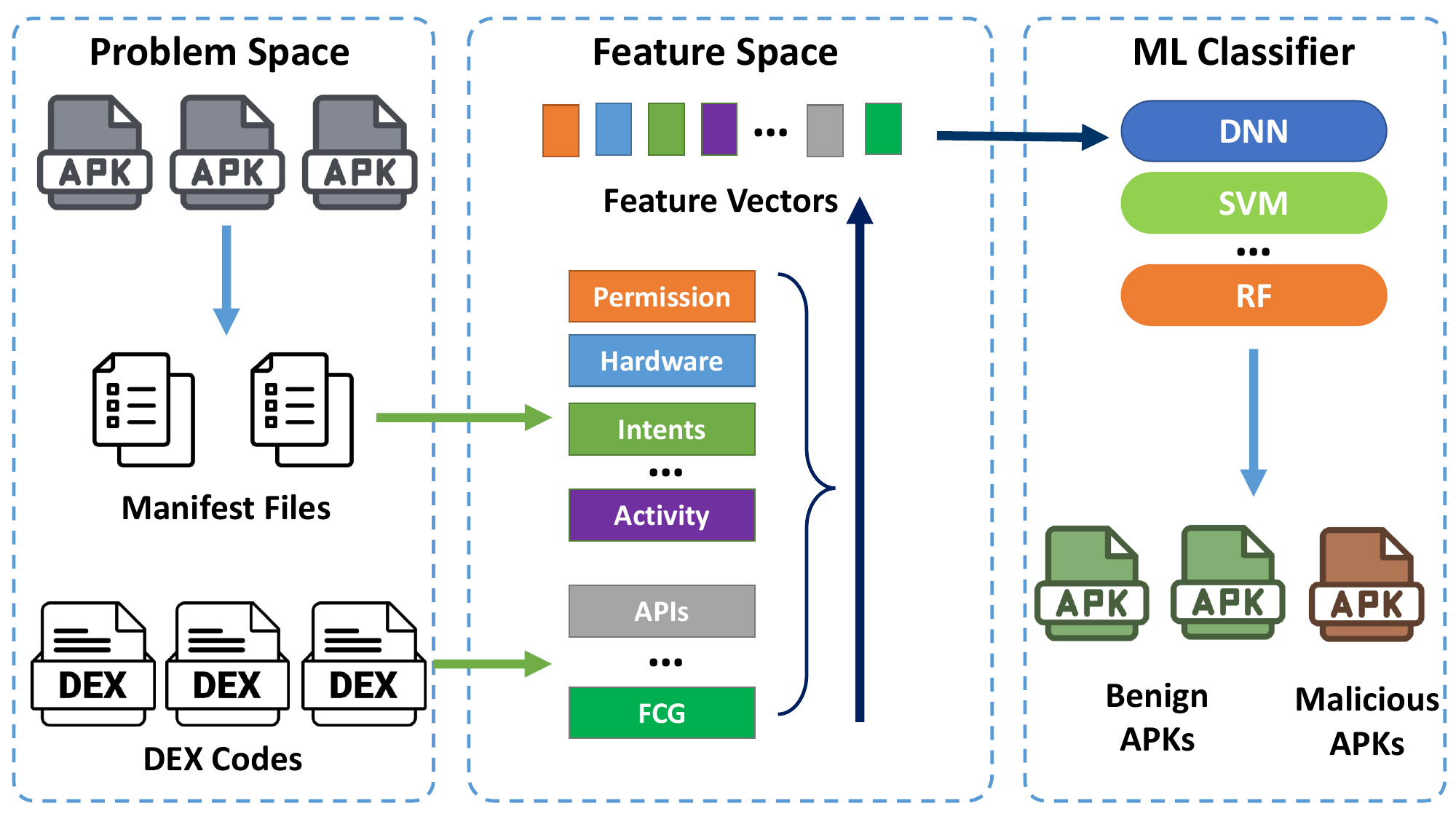}
	\caption{ML-based Android malware detection framework.}
	\label{fig:amd}
\end{figure}

\subsection{ML-based AMD}
\label{sec:MLAMD}

ML-based AMD methods are widely employed in real-world applications due to their efficiency and accuracy.
Typically, these methods contain two key stages: feature extraction and model prediction, as illustrated in Figure~\ref{fig:amd}.
The ML-based AMD methods first utilize program analysis tools~\cite{DBLP:conf/cascon/Vallee-RaiCGHLS99} to analyze the configuration information and the behavior of the applications.
The obtained analysis results are then used to extract specific features for the machine learning process.
The choice of features can vary significantly depending on the specific ML-based AMD method in use.
For instance, Drebin~\cite{DBLP:conf/ndss/ArpSHGR14} primarily considers the syntax features (static features), e.g., permissions, while MaMadroid~\cite{DBLP:conf/ndss/MaricontiOACRS17} only considers the semantic features (dynamic features), e.g., function call.
Once the features are extracted, they are organized into a feature vector.
This organized data is then used in the model prediction stage.
In this phase, ML-based AMD methods train machine learning classifiers to differentiate between benign and malicious applications.
Different ML-based AMD methods may employ different ML classifiers, e.g., Drebin~\cite{DBLP:conf/ndss/ArpSHGR14} utilize the Support Vector Machine (SVM) classifier and the MaMadroid~\cite{DBLP:conf/ndss/MaricontiOACRS17} can utilize the Random Forest (RF) classifier.

We utilize four SOTA and representative ML-based AMD methods in our evaluation.
They are Drebin~\cite{DBLP:conf/ndss/ArpSHGR14}, Drebin-DL~\cite{DBLP:conf/esorics/GrossePMBM17}, MaMadroid~\cite{DBLP:conf/ndss/MaricontiOACRS17}, and APIGraph~\cite{DBLP:conf/ccs/ZhangZZDCZZY20}.
The details about the four ML-based AMD methods can be found in Appendix~\ref{appendix:mlamd}.
The rationale behind selecting the four methods can be attributed to their well-acknowledged performance in malware detection and popularity in research~\cite{DBLP:conf/sp/PierazziPCC20,DBLP:journals/tifs/0002LWW0N0020,DBLP:journals/corr/abs-2303-08509,DBLP:conf/ccs/ZhaoZZZZLYYL21}.
For instance, Drebin~\cite{DBLP:conf/ndss/ArpSHGR14} achieved an AUROC score of 0.96 in our evaluation (Table~\ref{tab:AMDPerformance}).
Additionally, these ML-based AMD methods have been the target models for previous attack studies~\cite{DBLP:conf/sp/PierazziPCC20,DBLP:journals/tifs/0002LWW0N0020,DBLP:journals/corr/abs-2303-08509,DBLP:conf/ccs/ZhaoZZZZLYYL21}.
Moreover, the diversity of their feature spaces and classifiers enhances the comprehensiveness of our evaluation.

Despite the remarkable achievements of the aforementioned ML-based AMD methods, they remain vulnerable to adversarial examples.
Existing attack works demonstrate these vulnerabilities~\cite{DBLP:conf/ccs/ZhaoZZZZLYYL21,DBLP:journals/corr/abs-2303-08509,DBLP:journals/tifs/0002LWW0N0020,DBLP:journals/tifs/LiL20,DBLP:journals/corr/abs-2110-03301} but have their own limitations, e.g., require the knowledge about the target AMD methods, be limited to specific ML-based AMD methods, need a substantial number of target model queries, and/or not be robust to pre-processing.
More details can be found in Section~\ref{sec:relatedwork}.
The vulnerabilities of ML-based AMD methods stem from two areas: feature representation and bias in real-world data, which create evasion opportunities.
Even though these methods leverage a wide spectrum of features to model the malicious behavior of malware, the features captured may not precisely portray all nuances of such behavior.
This limitation provides a loophole for adversarial malware to evade detection.
On the other hand, in real-world deployments, machine learning classifiers may encounter biases stemming from spatial and temporal factors \cite{DBLP:conf/uss/PendleburyPJKC19,DBLP:conf/uss/Yang0HCAX021,DBLP:conf/sp/BarberoPPC22,DBLP:conf/uss/JordaneySDWPNC17,DBLP:conf/uss/ArpQPWPWCR22}.
These biases have the potential to negatively influence the decision boundaries of classification models, which in turn render these models suboptimal~\cite{DBLP:journals/corr/TanayG16,DBLP:conf/sp/Carlini017,li2023sok,YangImp2022,DBLP:conf/ccs/GuoMXSWX18,DBLP:conf/sp/ChenJW20,DBLP:journals/tdsc/LiJWLSBGWW22}.
Consequently, existing ML-based AMD methods face considerable threats from attacks, which exploit these vulnerabilities to undermine the accuracy and effectiveness of ML-based AMD methods.

\section{Methodology}

In this section, we first describe the threat model associated with \system.
Next, we provide an overview of the attack framework, which contains three stages from a practical perspective.
Finally, we elucidate the detailed methodology of \system by the three stages.

\begin{figure*}[t]   
	\centering  
	\includegraphics[width=0.75\linewidth]{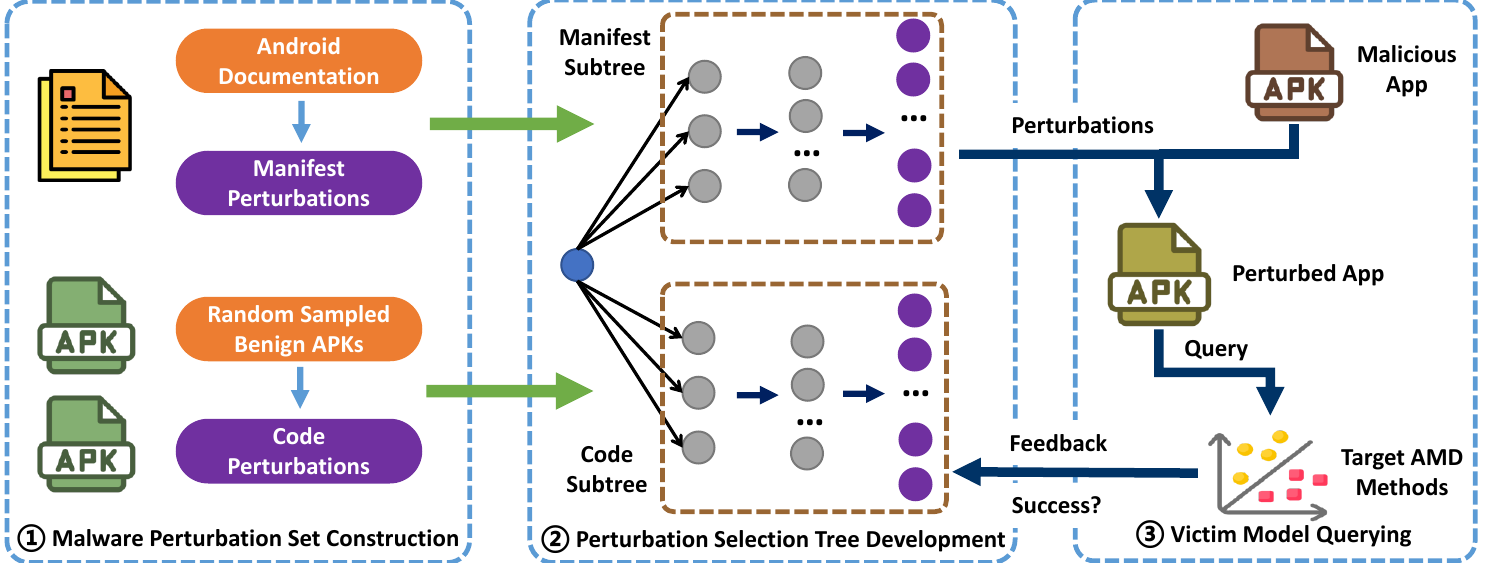}
	\caption{The overview of \system. It operates in three stages: malware perturbation set construction, perturbation selection tree development and victim model querying.}
	\label{fig:adz}
\end{figure*}

\subsection{Threat Model}
\label{sec:threatmodel}

We present the threat model by delineating three key components: adversary goals, knowledge and capabilities, building upon previous works~\cite{DBLP:journals/corr/abs-2212-14315,DBLP:journals/corr/abs-1902-06705,DBLP:conf/uss/SuciuMKDD18,DBLP:journals/pr/BiggioR18,DBLP:conf/sp/PierazziPCC20}.

\paragraphbe{Adversary Goals}
The adversary aims to execute misclassification attacks against a given ML-based AMD method.
This entails introducing a sequence of perturbations $\mathbf{P^{*}}$ within the perturbation space $ \mathbb{P} $ to a malware sample in the problem space, causing the target model $g$ to misclassify it as benign ($b$).
The adversary focuses solely on the misclassification of the malicious samples.
Consequently, the adversary goals can be formulated as:
\begin{equation}
\label{equ:goal}
    \mathbf{P^{*}} = \operatorname*{arg\,min}_{\mathbf{P} \in \mathbb{P}} \mathop{cost}(\mathbf{P}),
\end{equation}
\begin{align*}
    \text{s.t.} ~~g(\phi (\mathbf{X}+\mathbf{P^{*}})) = b,~ \mathop{malicious}(\mathbf{X}) = \mathop{malicious}(\mathbf{X} + \mathbf{P^{*}}),
\end{align*}
where $ \mathop{cost}() $ denotes the metric function employed to assess the cost of generating perturbations (e.g., human efforts, runtime overheads), $ \phi() $ represents the feature extraction function, $ \mathop{malicious}() $ signifies the malicious functionality verification function.
The adversary wants to find a sequence of perturbations with minimal cost.
The sequence of perturbations can make the target model misclassify the malware ($ g(\phi (\mathbf{X}+\mathbf{P^{*}})) = b $).
However, the sequence of perturbations can not influence the malicious functionality of the malware ($ \mathop{malicious}(\mathbf{X}) = \mathop{malicious}(\mathbf{X} + \mathbf{P^{*}}) $).

\paragraphbe{Adversary Knowledge}
Drawing from the zero knowledge setting defined in previous work~\cite{DBLP:conf/sp/PierazziPCC20}, the adversary lacks information about the target system, being only aware that static analysis is employed on the programs.
The adversary remains uninformed about feature spaces, model parameters, and the training dataset.
However, the adversary can acquire certain open-source knowledge, such as the Android documentation~\cite{AndroidDocumentation}, and access publicly available benign applications from app markets~\cite{GooglePlay} or app dataset~\cite{DBLP:conf/msr/AllixBKT16}.

\paragraphbe{Adversary Capabilities}
As more queries imply higher financial costs and an increased probability of detection, we assert that the adversary's ability to query the target system is limited.
The allocation of query budgets reflects the ability of the attacker, as the higher query budgets mean the stronger attacker.
We explore different query budgets in our experiments representing the attackers with different capabilities.
Additionally, we assume that the adversary can access the confidence provided by the target system.
This assumption is based on the fact that outputs from some real-world antivirus systems (e.g., VirusTotal) can be viewed as confidence, exemplified by the percentage of detected antivirus engines.
We acknowledge that in certain scenarios, the attacker may not have the capacity to inspect the output of some ML-based AMD methods.
We delve into a more detailed discussion about it in Section~\ref{sec:limitation}.

\subsection{Framework Overview}
\label{sec:overview}

At a high level, \system is a query-based attack framework to generate adversarial Android malware against ML-based AMD methods.
It operates through three stages: (1) malware perturbation set construction, (2) perturbation selection tree development, and (3) victim model querying, as shown in Figure~\ref{fig:adz}.
In the following, we illustrate every stage briefly, and the algorithmic descriptions for \system are provided in Appendix~\ref{appendix:algorithm}.

\paragraphbe{Malware Perturbation Set Construction}
Without knowledge of the target ML-based AMD method, \system incorporates two kinds of malware perturbation from open-source information: manifest perturbations and code perturbations.
The manifest perturbations targeting the syntax features (static features), e.g., permission, derive from Android documentation.
The code perturbations targeting the semantic features (dynamic features), e.g., function call, derived from 100 benign applications selected uniformly at random, absent from the training set of target ML-based AMD methods.

\paragraphbe{Perturbation Selection Tree Development}
\system builds the perturbation selection tree to guide the perturbation selection based on the semantic meanings inherent within the malware perturbations.
The tree's leaf nodes represent specific malware perturbations, while internal nodes symbolize shared semantics of the leaf nodes within the associated subtree.

\paragraphbe{Victim Model Querying}
\system selects malware perturbations iteratively using the perturbation selection tree, injects the selected perturbations, queries the target model with the perturbed application, and updates the perturbation selection tree based on the received feedback through a semantic-based adjustment policy. 
This iterative process continues until the framework achieves success or exhausts the query budget.

\subsubsection{Key Intuition}

Perturbations within an APK exhibit varying levels of evasion effectiveness.
Some perturbations have positive attack effectiveness by reducing the malware label's confidence, while others can impede such attacks by increasing it.
Furthermore, these perturbations exhibit significant semantic depth.
For example, the perturbations in the manifest file involve setting a target element with a specific value.
Adding a \textit{uses-feature} element with an \textit{android.hardware.audio.output} value relates to the declared hardware and software features required by the application.
In this case, the \textit{android.hardware.audio.output} value corresponds to hardware features, specifically audio requirements.

Upon understanding the semantics of perturbations, we posit that those with similar semantics are more likely to display comparable evasion effectiveness.
This is based on the observation that benign Android applications employ semantically similar elements in the manifest file to achieve specific functionalities.
We discuss the validation of the assumption in Appendix~\ref{appendix:validation}.
Consequently, once a perturbation is identified with positive attack effectiveness, we can infer other perturbations with positive attack effectiveness based on semantic meaning, thereby facilitating the selection of optimal perturbations.

Based on these insights, \system organizes perturbations in a tree structure according to their semantic depth, accommodating the heterogeneous perturbation space.
To reduce this space, \system groups semantically related perturbations and executes them simultaneously.
After receiving model feedback, \system adaptively adjusts the selection probabilities within the perturbation selection tree in a cascading manner, enhancing query efficiency and improving understanding of the target model.

\subsection{Malware Perturbation Set Construction}
\label{sec:mps}

As demonstrated in Section~\ref{sec:threatmodel}, the adversary lacks knowledge of the feature spaces, model parameters, and training dataset of the target model.
Nevertheless, the adversary can obtain open-source materials such as Android documentation and benign applications to derive the malware perturbation set.
Moreover, malware manipulation has three requirements stated by previous works~\cite{DBLP:journals/corr/abs-2303-08509}: all-feature influence, functional consistency, and robustness to pre-processing.
We discuss the detailed requirements and existing Android malware manipulation methods in Appendix~\ref{appendix:manipulation}.

To date, imposing all-feature influence under a zero knowledge setting remains inadequately explored.
To address this issue, the perturbation set is proposed encompassing both manipulations on \textit{AndroidManifest.xml} and manipulations on DEX code.
By employing these two types of perturbations, \system can mislead mainstream ML-based AMD methods (Section~\ref{sec:MLAMD}), regardless of whether they extract features, e.g., syntax features, semantic features, from \textit{AndroidManifest.xml} or DEX codes.
It is worth noting that if the feature knowledge of the targeted ML-based AMD methods is available, the malware perturbation set can be limited in order to perform more precise attacks.
For example, the malware perturbation set can only contain the code perturbations when targeting the function call graph-based AMD methods in practice.
However, since there is no constraint on the perturbation space, we consider broader perturbations to enhance the overall adaptability.

In accordance with the functional consistency requirement, any perturbation should not remove existing items from the \textit{AndroidManifest.xml} or DEX codes. However, it is feasible to add new items or codes which do not interfere with the original functionalities of the manipulated application.
Specifically, to manipulate \textit{AndroidManifest.xml}~\cite{AndoridManifestDocumentation}, we consider adding the \textit{uses-feature} element, the \textit{use-permission} element, and the \textit{action} element or \textit{category} element.
For the \textit{uses-feature} element, we take all referenced information about hardware features and software features into account~\cite{ReferencedFeatures}.
Regarding the \textit{uses-permission} element, the permission system is designed to protect security and privacy, which is a critical feature for identifying malware~\cite{DBLP:conf/ccs/AuZHL12}.
Every permission has a protection level that characterizes the potential risk implied by the permission~\cite{AndoridPermission}.
We argue that if a malicious application possesses permissions at a dangerous level, it is more likely to be detected.
Consequently, we only add permissions with normal or signature protection levels.
Concerning \textit{action} and \textit{category} elements, these elements, encapsulated in the \textit{intent-filter} element, provide information about the intents of the application.
However, adding them directly to existing activities or services might interfere with the original functionality.
Therefore, we create new \textit{activity} elements and \textit{receiver} elements for these elements.
To preserve functionality, the newly added \textit{activity} elements and \textit{receiver} elements should not be launched during the application's lifetime.
Their names are randomly generated to prevent explicit intents from the original code from launching them.
Furthermore, to avoid being launched by implicit intents, the \textit{action} and \textit{data} elements with the random URI are added to the \textit{activity} and \textit{receiver}.
In terms of added values, we consider the standard activity actions, standard broadcast actions, and standard categories in Android documentation~\cite{AndroidIntent} because they are established norms.
Specific examples are provided in Appendix~\ref{appendix:manifest}.

\begin{figure}[t]   
	\centering  
	\includegraphics[width=1\linewidth]{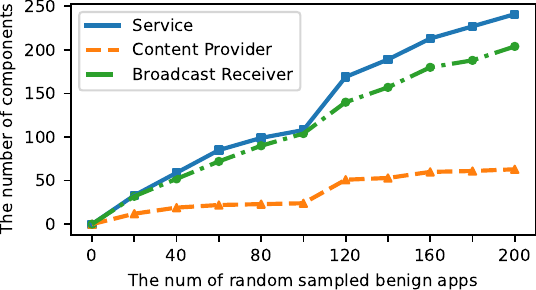}
	\caption{{The number of services, content providers and broadcast receivers with the number of randomly sampled benign applications.}}
	\label{fig:num_components}
\end{figure}

To manipulate the DEX codes, \system adds three types of Android app components from benign applications, i.e., service, broadcast receiver, and content provider~\cite{AppComponents}.
These components are selected for the following reasons.
On the one hand, these components can be easily obtained from benign applications.
We randomly sample benign applications from AndroZoo, then tally the numbers of services, broadcast receivers, and content providers according to their \textit{AndroidManifest.xml}.
As depicted in Figure~\ref{fig:num_components}, only 100 sampled benign applications yield 108 different services, 24 different content providers, and 104 different broadcast receivers.
This indicates that these components are abundant in benign applications.
On the other hand, these components possess rich semantics, which is beneficial for reducing queries.
We enumerate classes and functions for our randomly chosen components from 100 benign applications.
The results show approximately 175 classes and 873 functions per service, 136 classes and 703 functions per broadcast receiver, and 417 classes and 2,044 functions per content provider.
With such abundant code information, adding components to malware significantly alters its semantic features, making it more likely to deceive the classifier.
Therefore, the iterative addition of components can quickly change the features of malware, leading to lower queries.

To maintain the functional consistency requirement, there are two cases for added components.
On the one hand, the added components should not be launched during APK execution.
Here, since no extra code is executed, the original functionality remains unaffected.
On the other hand, if the added components are launched, their execution should not affect the original part of the malware code.
Even though these components execute, they act as independent entities without impacting the original functionality.

To satisfy the first case, \system does not inject explicit invokes in the original part of the code as done in previous work~\cite{DBLP:conf/sp/PierazziPCC20}.
However, simply adding components to malware is easily identifiable, as the added code can be viewed as an isolated part of the code, which static analysis tools can effortlessly remove.
This violates the requirement for robustness to pre-processing.
To achieve robustness in pre-processing, we leverage Android's intent mechanism.
Specifically, when registering these components in \textit{AndroidManifest.xml}, \system sets the attributes of \textit{android:exported} and \textit{android:enabled} to \textit{true}.
By setting \textit{android:enabled} to true, the component is enabled and can be started and respond to intents, indicating that it is not dead code.
By setting \textit{android:exported} to true, the component can be accessed or invoked by other applications on the device, not just within the application that defines it.
This allows external applications to interact with the component, provided they have the necessary permissions, ensuring that the added components cannot be considered dead code.

To satisfy the second case, we ensure the original malware functionality remains unaffected even if other applications invoke the added components.
This is achieved using Android's subprocess mechanism~\cite{AndroidProcess}.
By setting the \textit{android:process} attribute with a random string, the added components run in another process, isolated at the system level from the original part of the malware.
Furthermore, the added components are randomly chosen from benign applications and have no inter-process communication with the original part of the malware.
Therefore, even if the added components are launched, the functionality of the malware is not influenced.
Activities can also be executed in the subprocess.
However, when an activity is invoked, it will be displayed on top of the current activity.
As a result, the user will see the new activity on the screen, and it influences the original functionality of the malware.
Hence, we exclude activities in our approach.
We provide examples in Appendix \ref{appendix:manifest}.

\begin{figure}[t]   
	\centering  
	\includegraphics[width=1\linewidth]{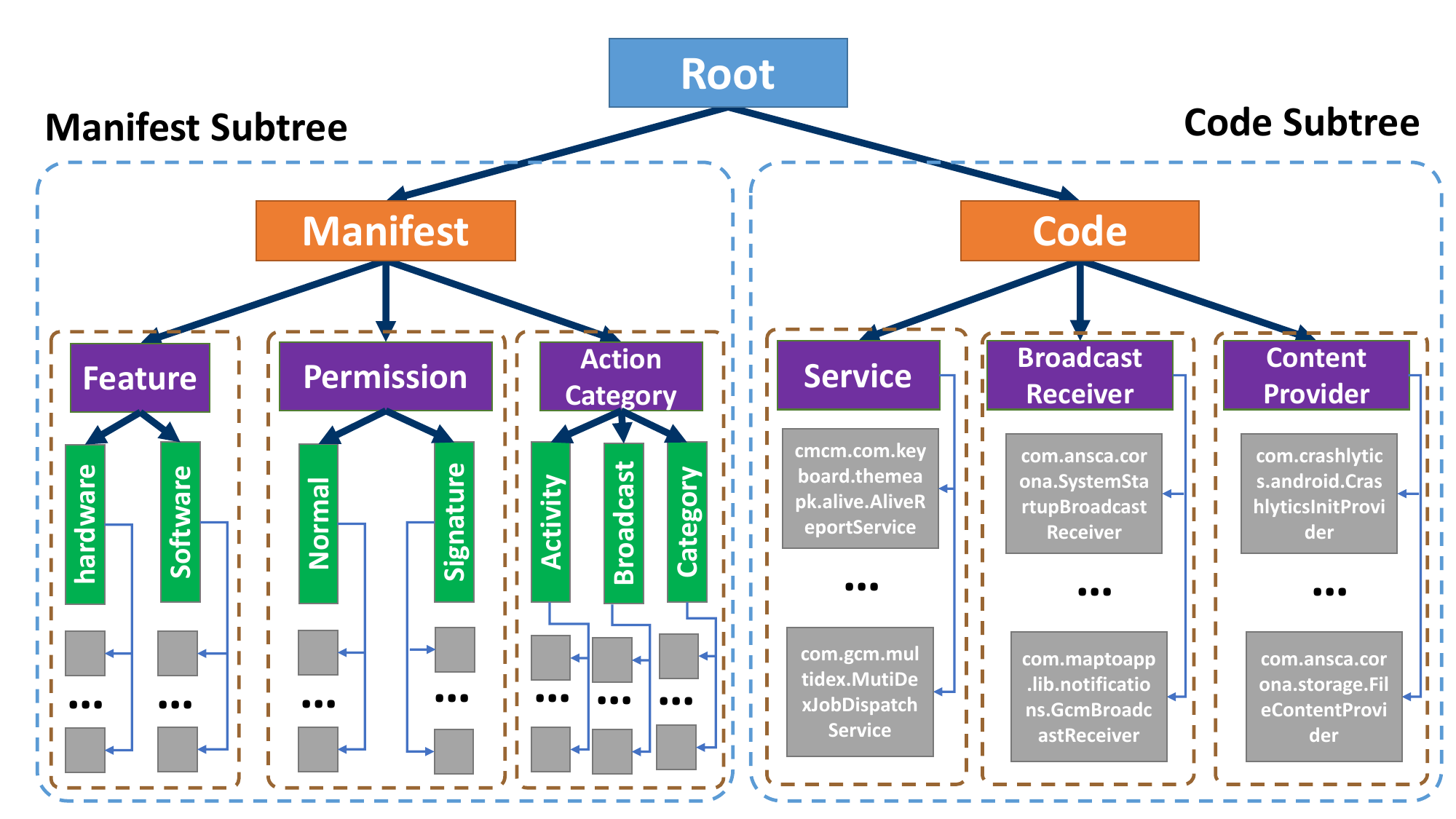}
	\caption{{The structure of perturbation selection tree.}}
	\label{fig:PST}
\end{figure}

\subsection{Perturbation Selection Tree Development}

Upon acquiring the malware perturbation selection set, it is crucial to select specific perturbations from the set.
However, this set contains numerous heterogeneous perturbations.
With 492 specific perturbations in our set, it becomes challenging to select an optimal perturbation sequence.
Moreover, the heterogeneous nature of the perturbation set makes it unsuitable for using an optimization algorithm to obtain the best solution \cite{DBLP:conf/ccs/ZhaoZZZZLYYL21}.
To address these challenges, \system employs a novel data structure called the perturbation selection tree, designed to select optimal perturbations.
The perturbation selection tree is designed with the following considerations in mind.
Firstly, akin to the decision tree algorithms~\cite{myles2004introduction,DBLP:journals/csur/Quinlan96} in machine learning that can process different feature types, such as numerical and categorical features, the perturbation selection tree model can handle heterogeneous perturbations using its tree structure.
Secondly, the semantic depth of perturbations is helpful for designing the tree structure.
By leveraging the unique design of the perturbation selection tree, \system can uncover hidden relationships between perturbations.
This enables it to extract more information from a single query, thereby reducing the number of queries required.

At a high level, the perturbation selection tree is a hierarchical tree structure in which leaf nodes represent specific perturbations.
During each selection procedure, \system samples a path from the root node to a leaf node, corresponding to the particular perturbations.
Internal nodes represent the specific semantics shared by the leaf nodes belonging to the subtree with the internal node as its root.
Figure \ref{fig:PST} illustrates the structure of the perturbation selection tree, which consists of two nodes at the first layer: the manifest node and the code node.

The manifest subtree represents the subtree organizing the perturbations related to modifications in the \textit{AndroidManifest.xml}.
\system firstly considers the element-level semantic.
Specifically, \system separates the perturbations into use-feature subtree, permission subtree, and action\&category subtree.
The use-feature subtree encompasses perturbations involving the addition of the \textit{uses-feature} element in the \textit{AndroidManifest.xml}, the permission subtree collects perturbations related to adding the \textit{permission} element, and the action\&category subtree comprises perturbations about adding the \textit{action} or \textit{category} elements.
\system then examines the semantics at the name level.
Within the use-feature subtree, element names follow the format \textit{android.hardware.xxx} or \textit{android.software.xxx}, where \textit{xxx} denotes a specific feature name, such as \textit{microphone} and \textit{bluetooth}.
Based on this observation, \system organizes the hardware and software subtree within the use-feature subtree.
In the permission subtree, permissions have different protection levels, prompting \system to categorize perturbations into normal and signature level groups.
Within the action\&category subtree, standard activity actions, standard broadcast actions, and standard category are considered, resulting in the formation of activity action, broadcast, and category subtree, respectively.

In the leaf node layer of the manifest subtree, specific perturbations still exhibit rich semantics derived from their text representation.
Consequently, \system clusters semantically similar perturbations into a single leaf node in the perturbation selection tree.
This approach enables \system to sample multiple perturbations simultaneously and execute them collectively, effectively reducing query costs.
To cluster perturbations in the uses-feature subtree is relatively straightforward due to the simple and regular text representations of the \textit{uses-feature} element names.
For instance, \textit{\seqsplit{android.hardware.audio.output}} and \textit{android.hardware.audio.pro} exhibit similar semantics, as they share the keyword \textit{audio} in the third position from left to right.
Therefore, \system clusters these perturbations based on the keyword.
However, for perturbations in the permission subtree and action\&category subtree, no explicit keywords are present in their name representations, rendering the clustering algorithm used in the uses-feature subtree unsuitable.
Instead, \system devises a customized clustering algorithm to cluster perturbations in these two subtrees based on their text representation.
For example, in the permission perturbation \textit{android.permission.ACCESS\_WIFI\_STATE}, three keywords (i.e., \textit{ACCESS}, \textit{WIFI}, and \textit{STATE}) can be extracted.
The algorithm iteratively clusters perturbation groups based on the similarity of the keywords.
The detailed clustering algorithm can be found in Appendix \ref{appendix:cluster}.

The code subtree represents the subtree in which the root node organizes perturbations related to modifying DEX codes. 
As demonstrated in Section~\ref{sec:mps}, \system injects three types of Android app components from benign applications into malware. 
These component types serve as explicit semantics for the perturbations. 
Consequently, \system constructs the service subtree, broadcast receiver subtree, and content provider subtree to accommodate these types.
Although various kinds of program semantics exist, such as function call graphs \cite{DBLP:journals/tosem/MurphyNGL98,DBLP:conf/icse/SalisSLSM21}, Android app components are relatively independent and capable of representing specific functions.
As a result, \system does not extract fine-grained semantics or cluster components together.
Instead, each leaf node in the code subtree represents a single Android app component, preserving the independence and functionality representation of these elements.

Upon constructing the structure of the perturbation selection tree, \system initializes the probabilities of each node in the perturbation selection tree in a bottom-up manner.
For leaf nodes in the manifest subtree, they comprise perturbation groups that combine multiple specific perturbations.
Consequently, larger perturbation groups are more likely to cause large changes in model confidence.
However, it is uncertain whether these confidence changes will benefit the attack (decreasing the model's confidence in the malware label) or harm it.
As such, these perturbations are risky; if a perturbation is detrimental to the attack, it may cause significant damage.
Conversely, smaller perturbation groups will have minimal impact on model confidence.
Considering these intuitions, it is advantageous to select perturbation groups with median sizes.
As a result, \system leverages a normal distribution to fit the sizes of perturbation groups and allocates initial probabilities based on the normal distribution.
For leaf nodes in the code subtree, \system assigns equal probabilities to them due to their independent properties.
For all internal nodes except for the manifest node and code node, \system aims to sample each leaf node uniformly.
Thus, \system assigns node probabilities based on the number of leaf nodes that are descendants of the node, with a larger number of leaf nodes corresponding to smaller selection probabilities.
The more number of leaf nodes, the smaller chosen probability.
Regarding the probabilities of the manifest node and code node, the default value is set to 0.5, as the target model is unknown.
However, if the attacker possesses prior knowledge about the feature type of the target classifier, they can adjust these values to expedite the attack process.

Lastly, in the perturbation selection process, \system samples a path from the root node to the leaf node.
Specifically, \system iteratively selects nodes at random based on their selection probabilities until a leaf node is chosen.

\begin{algorithm}[t]
    \footnotesize
    \caption{Semantic-based Adjustment Policy}
    \label{alg:adjust}
    \begin{algorithmic}[1]
        
        \Require{
        Perturbation selection tree $\mathbb{T}$;
        Perturbation node $P$;
        Previous malicious confidence $y$;
        Malicious confidence $y'$.
        }
        
        \Ensure{
        Adjusted perturbation selection tree $\mathbb{T'}$.}
        
        \Statex
        \While {True} \Comment{Recursively delete the perturbation node $P$.}
            \State $P' \leftarrow $ GetParentNode($P$) \Comment{Get the parent perturbation node $P'$.}
            \If {$P'$ has other children}
                \State TransferProba($P$, $P'$) \Comment{Transfer probability of P to other children.}
                \State \textbf{break}
            \Else 
                \State $P \leftarrow P'$
            \EndIf
        \EndWhile

        \If {$y' \geq y$}   \Comment{Adjust the internal node probability.}
            \State $P \leftarrow P'$
            \State $P' \leftarrow $ GetParentNode($P'$)
            \While{$P'$ is not root} 
                \State ReInitProb($P'$) \Comment{Reinitialize the internal node probability.}
                \If {$y' = y $} \Comment{Add penalty when no effectiveness.}
                    \State AddPenalty($P$, $P'$)
                \EndIf
                \State $P \leftarrow P'$
                \State $P' \leftarrow $ GetParentNode($P$)
            \EndWhile
        \EndIf
        
        \If {$y' \geq y$} \Comment{Adjust the first layer probability.}
            \State AdjustFirstLayer($\mathbb{T}$)
        \EndIf
        \State \Return $\mathbb{T}'$
    \end{algorithmic}
\end{algorithm}

\subsection{Victim Model Querying}

After developing the perturbation selection tree, \system then proceeds with a four-step iterative process: selecting the malware perturbation, perturbing the malicious applications, obtaining the model feedback, and updating the perturbation selection tree.
Initially, \system selects the malware perturbation from the tree by sampling a path from the root to a leaf node.
Then, \system uses program analysis tools, such as FlowDroid~\cite{DBLP:conf/pldi/ArztRFBBKTOM14}, to implement the chosen perturbations to the malicious application.
After this, the perturbed application is submitted to the target model to garner feedback.
Finally, \system refines the perturbation selection by using a semantic-based adjustment policy leveraging the received model feedback.
The iterative process continues until either success is achieved or the query budget depletes.

In the victim model querying step, the semantic-based adjustment policy is a crucial element, as it allows the perturbation selection tree to learn from the target model adaptively.
The underlying intuition for this semantic-based adjustment policy is that if perturbations positively impact malware evasion (decreasing the model's confidence in the malware label), the probability of semantically related perturbations should be increased.
Conversely, if perturbations negatively impact malware evasion, the probability of semantically related perturbations should be decreased.
If perturbations have no effect on malware evasion (no changes in the model's confidence in the malware label), it means that the target model may not use the features altered by the perturbation.
Then the probability of semantically related perturbations should be decreased with a penalty.
Consequently, \system imposes an additional penalty to reduce the probability of all nodes that are ancestors of the perturbation node.

The details of the adjustment policy algorithm can be found in Algorithm \ref{alg:adjust}.
Upon obtaining the selected perturbation node, the adjustment policy removes it and equally distributes its probability among its sibling nodes (lines 1-9).
After this deletion, the internal nodes with the selected perturbation node as a descendant should have their probability decreased.
However, if the selected perturbation node has a positive impact, the probability of these nodes should increase, which is achieved by not decreasing the probability.
Otherwise, the probability is reduced by computing it based on the leaf node count used during the initialization phase (lines 10-21).
If the selected perturbation node has no impact, an extra penalty is applied to its ancestor nodes (lines 15-17).
The penalty percentage is calculated as the node depth multiplied by a constant (0.1 in our experiments), resulting in deeper nodes receiving larger penalties.
Lastly, the probability of the first-layer nodes is adjusted independently (lines 22-24) to identify the target model's feature type quickly.
Consequently, the corresponding manifest node or code node probability is halved if the selected perturbation node does not positively affect evasion.

\section{Evaluation}

In this section, we conduct comprehensive experiments to evaluate the performance of \system.
We first present the experimental settings, including the datasets, target AMD methods, and evaluation metrics.
We then assess the effectiveness and cost of \system.
Subsequently, we examine its performance in the context of real-world antivirus solutions.
Furthermore, we explore the effectiveness of \system against the dynamic analysis-based defense.
Finally, we evaluate the functionality consistency requirement by both static and dynamic analysis.

\subsection{Experimental Setup}

\paragraphbe{Implementation Details}
\system is an automatic attack framework with three stages to generate adversarial Android malware.
Our implementation of the prototype of \system utilizes a hybrid approach combining both Python and Java.
We use Python to handle the program logic of the attack process, encompassing the data structure of the malware perturbation set, the development of the perturbation selection tree and the semantic-based adjustment policy, as well as managing the iterative query process during the victim model querying stage.
On the other hand, Java is utilized for the extraction and implementation of perturbations.
In order to slice Android app components to construct the malware perturbation set and perform program manipulation in the stage of victim model querying, the framework integrates FlowDroid~\cite{DBLP:conf/pldi/ArztRFBBKTOM14}, which is built upon Soot~\cite{DBLP:conf/cascon/Vallee-RaiCGHLS99}.

In practice, our implementation automatically selects the malware perturbations (e.g., permissions) from the perturbation selection tree.
It then automatically employs FlowDroid to modify the application by incorporating these chosen perturbations.
It continues by autonomously querying the targeted ML-based AMD method with the modified application.
The feedback from this query helps refine the perturbation selection tree.
This iterative process continues until either success is achieved or the query budget depletes.
To facilitate further research, the code and data of \system are responsibly shared with other researchers upon request\footnote{The instructions regarding access requests can be found at: \url{https://github.com/gnipping/AdvDroidZero-Access-Instructions}.} following previous works~\cite{DBLP:conf/sp/PierazziPCC20,DBLP:conf/ccs/ZhaoZZZZLYYL21} due to potential ethical concerns (Section~\ref{sec:ethic}).

\begin{table*}[t]\small\centering
	\setlength{\abovecaptionskip}{0pt}
	\caption{Attack performance of \system and baseline method measured by ASR. It can be seen that \system outperforms all baseline methods under most settings of query budget, target AMD methods and ML classifiers.}
 
	\begin{tabular}{@{}C{2.5cm}C{1.8cm}C{1.7cm}C{1.7cm}C{1.7cm}C{1.7cm}C{1.7cm}@{}}
		\toprule
		\multirow{5}{*}{\begin{tabular}[c]{@{}c@{}}Attack Methods\end{tabular}} &
		\multirow{5}{*}{Query Budgets}                                                           &
		\multicolumn{5}{c}{Target AMD Methods}                                                                                                                                                                                                \\ \cmidrule(l){3-7}
		                                                                                   &            &    Drebin      &   Drebin-DL   &    APIGraph      &  \multicolumn{2}{c}{MaMadroid}                                                        \\ \cmidrule(l){3-7}
		                                                                                   &
		                                                                                   &
		\begin{tabular}[c]{@{}c@{}}SVM\end{tabular}                            &
		\begin{tabular}[c]{@{}c@{}}MLP\end{tabular}                         &
		\begin{tabular}[c]{@{}c@{}}SVM\end{tabular}                           &
		\begin{tabular}[c]{@{}c@{}}RF\end{tabular}                            &
		\begin{tabular}[c]{@{}c@{}}3-NN\end{tabular}                                  \\ \midrule
		\multirow{4}{*}{\system}                                              & 10       & 57\%                      & 67\%                      & 50\% & 92\% & 83\%  \\
		                                                                                   & 20  & 80\%                      & 80\%                      & 70\% & 98\% & 93\%  \\
		                                                                                   & 30      & 87\%                      & 85\%                               & 94\%          & 100\% & 100\%   \\
		                                                                                   & 40           & 100\%                      & 90\%                               & 94\% & 100\% & 99\%  \\
		                                                                                   \midrule
		\multirow{4}{*}{MAB}                                                            & 10       & 48\%                               & 41\%                               & 31\%          & 91\%          & 84\%           \\
		                                                                                   & 20  & 63\%                              & 78\%                               & 60\%          & 98\%          & 97\%           \\
		                                                                                   & 30     & 78\%                               & 84\%                              & 77\%          & 100\%         & 98\%          \\
		                                                                                   & 40        & 98\%                             & 91\%                              & 83\%          & 100\%         & 100\%               \\
		                                                                                   \midrule
		\multirow{4}{*}{RA}                                                               & 10       & 39\%                               & 44\%                              & 35\%          & 93\%         & 92\%             \\
		                                                                                   & 20  & 62\%                               & 77\%                              & 58\%          & 98\%         & 90\%                   \\
		                                                                                   & 30      & 85\%                               & 80\%                              & 76\%         & 100\%         & 86\%                \\
		                                                                                   & 40      & 90\%                               & 88\%                            & 87\%         & 100\%          & 98\%                \\
		                                                                                   \bottomrule
	\end{tabular}
 
	\label{tab:attackeffect}
\end{table*}

\paragraphbe{Dataset}
Our primary dataset comprises 135,859 benign applications and 15,778 malware samples, totaling 151,637 applications.
This dataset is derived from the previous work \cite{DBLP:conf/sp/PierazziPCC20}.
We download the APKs from AndroZoo \cite{DBLP:conf/msr/AllixBKT16} based on the SHA-256 value provided in the aforementioned study \cite{DBLP:conf/sp/PierazziPCC20}.
As a result, we successfully obtain 151,637 applications dated between January 2016 and December 2018.
The dataset has already been processed by Pierazzi \textit{et al}. \cite{DBLP:conf/sp/PierazziPCC20}, adhering to the labeling criteria outlined in Tesseract~\cite{DBLP:conf/uss/PendleburyPJKC19}.

However, we employ a time-aware split~\cite{DBLP:conf/uss/PendleburyPJKC19,DBLP:conf/uss/ArpQPWPWCR22} for our dataset, which differs from the approach taken in the previous work~\cite{DBLP:conf/sp/PierazziPCC20}.
Our work aims to generate adversarial Android malware under the zero knowledge setting, a highly practical scenario.
In contrast, Pierazzi \textit{et al}. sought to reveal the weaknesses of machine learning classifiers in the perfect knowledge context.
In practical applications, the concept drift problem naturally exists and must be considered.
Therefore, we perform a time-aware split of the dataset to simulate a \textit{real malware classifier}.
Specifically, we utilize applications dated between January 2016 and December 2017 as the training set and applications dated between January 2018 and December 2018 as the test set.

In order to incorporate more recent malware samples, we also employed a supplementary dataset provided by VirusShare~\cite{virusshare}, consisting of 20,206 malware samples from 2022.
The rapidly evolving landscape of Android APKs and the fact that the most recent samples in our primary dataset are only from 2018 make this supplementary dataset crucial for assessing the effectiveness of \system in a temporal context.
Specifically, we evaluate \system against VirusTotal using the supplementary dataset (Section~\ref{sec:timevalidity}).

\paragraphbe{Target Model}
We select four SOTA ML-based AMD methods, namely Drebin \cite{DBLP:conf/ndss/ArpSHGR14}, Drebin-DL \cite{DBLP:conf/esorics/GrossePMBM17}, MaMadroid \cite{DBLP:conf/ndss/MaricontiOACRS17}, and APIGraph \cite{DBLP:conf/ccs/ZhangZZDCZZY20}, as our target ML-based AMD methods.
To ensure fidelity to their original implementations, we strictly adhere to the descriptions and configurations provided in their respective publications.
Consequently, for Drebin and APIGraph, we employ SVM with the linear kernel as the target classifier, while for Drebin-DL, we utilize a two-layer multi layer perceptron (MLP) as the target classifier.
For MaMadroid, we use RF and 3-Nearest Neighbor (3-NN) as target classifiers.
These target models incorporate various feature types and machine learning classifiers, thereby offering a diverse representation of ML-based AMD methods. 
Further details regarding the implementation and the detection performance of these target models can be found in Appendix \ref{appendix:targetmodel}.

\paragraphbe{Metric}
In our experimental evaluation, we conduct attacks using true positive malware from the test set in the primary dataset, meaning that we target malware samples that have been accurately classified as malware.
Intuitively, there is no need to generate adversarial examples for malware applications already misclassified as benign since they do not require any queries.
Consequently, we employ the attack success rate (ASR) as the evaluation metric for the effectiveness of \system.
ASR represents the proportion of successfully generated adversarial Android malware instances (denoted by $N_{s}$) to the total number of malware samples utilized for the attack (denoted by $N_{t}$), i.e., ASR = $N_{s} / N_{t}$.

Additionally, in order to measure the attack cost associated with \system, we take into account both human factors and runtime overheads, as recommended by Apruzzese \textit{et al}.~\cite{DBLP:journals/corr/abs-2212-14315}.
From the perspective of human factors, we employ the duration incorporating the attack preparation and design phases, which includes tasks such as source code writing to measure the attack cost.
To measure the attack cost from the perspective of runtime overheads, we assess the program execution time and the number of query times (QT).
The program execution time refers to the time from the query beginning to the generation of adversarial Android malware serving as a reflection of the CPU cost to some extent, whereby a shorter duration indicates a more scalable attack.
Meanwhile, QT signifies the number of queries executed in the process of generating a single adversarial Android malware instance.
It provides insights into the potential financial cost and detection likelihood, considering the pricing strategies of commercial machine learning models and the established correlation between increased queries and the detection probability~\cite{DBLP:conf/uss/LiSWZ0Z22}.

\paragraphbe{Experimental Environment}
We run all experiments on a Ubuntu 20.04 server with 251G memory and 39G swap memory, 2 Intel(R) Xeon(R) Gold 6346 CPUs and one NVIDIA RTX 3090 GPU.

\subsection{Attack Performance}

To assess the performance of \system, we apply it to attack the aforementioned mainstream target models.
The evaluation of the attack performance encompasses two dimensions: attack effectiveness and attack cost.
Moreover, we compare \system with two attack methods, namely MAB \cite{DBLP:conf/asiaccs/SongLAGKY22} and Random Attack (RA).
Detailed information regarding the baseline methods can be found in Appendix \ref{appendix:baseline}.

\subsubsection{Attack Effectiveness}
To evaluate attack effectiveness, we establish different query budgets for generating adversarial Android malware.
We consider the generation of adversarial Android malware successful if the attack algorithm can generate it within the query budget.
To be specific, we randomly select 100 true positive malware samples from the test set within our primary dataset to carry out the attack, implying that the target malware samples have been correctly classified as malware.
Subsequently, we compute the ASR in every case, as shown in Table~\ref{tab:attackeffect}.
We also report the actual number of successfully generated adversarial Android malware instances in every case, which can be found in Appendix~\ref{appendix:actualnumber}.

Table \ref{tab:attackeffect} presents the attack effectiveness of \system and baseline methods against mainstream ML-based AMD methods using static program analysis approaches under varying query budgets.
\system outperforms the baseline methods in most settings of query budget, target AMD methods, and ML classifiers.
Additionally, \system achieves nearly 100\% ASR against all AMD methods, regardless of feature type or ML classifier, when the query budget is set to 40.
This demonstrates that \system exhibits strong attack effectiveness against ML-based AMD methods using static program analysis under zero knowledge settings.

\system performs better when the target model only considers graph-based features.
For example, as shown in Table~\ref{tab:attackeffect}, the ASR of \system is 92\% for the MaMadroid method with the RF classifier when the query budget is set to 10.
However, the ASR of \system drops to 57\% when the target AMD method is Drebin with the SVM classifier under the same query budget.
This can be attributed to two factors.
First, the adjustment policy design in \system allows for quickly learning the feature type of the target AMD methods.
The adjustment policy modifies the probability of manifest nodes and code nodes exponentially, enabling \system to choose effective perturbations with just a few queries.
Second, the Android app component-level perturbations are effective due to the rich program semantics within these components.
The rich program semantics can significantly change the function call graph, resulting in substantial changes to the malware's feature value.
Furthermore, since these Android app components are sourced from benign applications, they intuitively have a positive impact on the attack.

Another observation is that the APIGraph-enhanced Drebin method is more robust than the original Drebin method.
For instance, under the 20 query budget setting, \system achieves a 70\% ASR on the APIGraph-enhanced Drebin method, while the ASR is 80\% for the original Drebin method.
This can be explained by the fact that APIGraph replaces individual functions with their corresponding clusters, providing a higher level of abstraction and reducing the impact of minor variations.
Consequently, during the ML model training process, ML models more readily learn robust features from APIGraph-enhanced features.

We also find that the 3-NN classifier is more robust than the RF classifier.
For example, under the 10 query budget setting, \system achieves a 83\% ASR for the MaMadroid method with the RF classifier, whereas the ASR is 92\% for the 3-NN classifier.
This conclusion holds for all baseline methods.
As noted by Li \textit{et al}. \cite{DBLP:journals/corr/abs-2303-08509}, this is because the 3-NN classifier takes into account the data samples in the training set when classifying a sample.
This process typically considers more features, making it easier to distinguish between benign applications and malware.

\begin{figure}[t]   
	\centering  
	\includegraphics[width=1\linewidth]{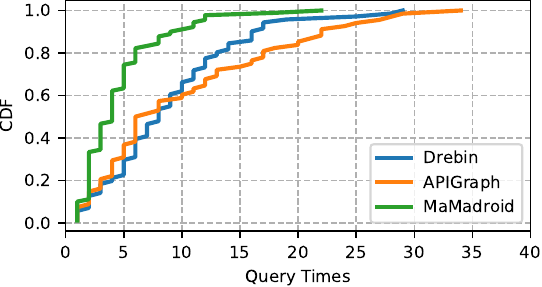}
	\caption{The CDF of query times in successfully evaded malware.}
	\label{fig:query_time}
\end{figure}

The performance of \system increases as the number of queries increases, as shown in Table~\ref{tab:attackeffect}.
For example, \system achieves 50\% ASR against APIGraph with 10 query budgets while 94\% ASR with 30 query budgets. 
However, with more queries, the potential financial cost and the detection probability increase as well.
Thus, we find that the 30-query budget is the optimal value to trade off the cost and utility because it is the smallest value to achieve 100\% ASR against the MaMadroid methods.
It is worth noting that \system can employ any query budget to conduct the attack depending on the practical capability of the attacker.

Compared to \system, all baseline methods are less effective against ML-based AMD methods, especially when the query budget is low.
For instance, \system achieves a 94\% ASR against APIGraph within 30 queries, a result unattainable by MAB and RA even with 40 queries.
The reason behind this observation is as follows: MAB does not concern the context of the perturbations and only accounts for the perturbation type.
Consequently, when MAB faces an AMD method like Drebin, it is less effective because the context information is crucial.
For FCG-based methods like MaMadroid, due to the effectiveness of Android app component-level perturbation, context information is less important, allowing it to achieve comparable attack effectiveness with \system.
RA does not consider any information in perturbation; therefore, its effectiveness depends on the perturbation set. 
The effectiveness of RA can be viewed as a result of our perturbation set's contribution.

In summary, \system demonstrates strong attack effectiveness against ML-based AMD methods using static program analysis approaches under zero knowledge setting.
The results highlight the advantages of the semantic-based adjustment policy design and Android app component-level perturbations. 
Additionally, the comparisons with baseline methods, such as MAB and RA, further emphasize the superiority of \system in handling a variety of target models with different feature types and ML classifiers.

\begin{figure}[t]   
	\centering  
	\includegraphics[width=1\linewidth]{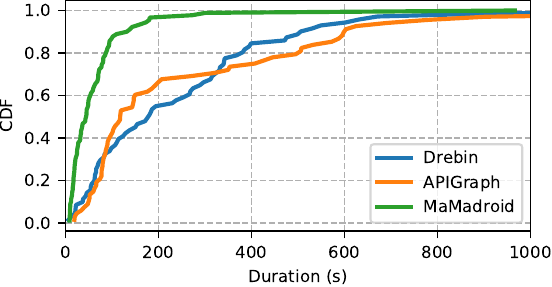}
	\caption{The CDF of the program execution time in successfully evaded malware.}
	\label{fig:query_duration}
\end{figure} 

\subsubsection{Attack Cost}
To evaluate the attack cost of \system, we examine two aspects: human factors and runtime overheads.
The human factors mainly involve the stages of malware perturbation set construction and perturbation selection tree development.
The two stages require extensive domain expertise and OSINT operations, including detailed analysis of Android documentation.
Based on our code update records, the two stages collectively take us approximately three months, highlighting their importance to our contribution.
Breaking it down further, we spent about a week identifying the malware perturbation set and preparing all malware perturbations.
The remainder of the time is devoted to designing the perturbation selection tree and the semantic-based adjustment policy.
However, other potential attackers who might use \system may only need to occasionally revisit the malware perturbation set construction stage, which requires human involvement, such as analyzing the latest Android documentation.
This stage requires considerably less time (one week for us) compared to designing the entire attack framework.

Once the initial setup is complete, the emphasis then shifts to the query process.
At this stage, the QT and the program execution time become significant metrics.
QT is important as it reflects the financial implications and detection likelihood of the attack, while the program execution time signifies the scalability of the attack.
We apply \system to attack against Drebin, APIGraph, and MaMadroid using RF, setting the query budget to 40 in 100 random sampled malware, which can be identified by these methods in the test set of the primary dataset.
In total, the numbers of successfully evaded malware are 74, 71, and 98, respectively.
We then record the QT and program execution time for all successfully evaded malware samples and plot the cumulative distribution function (CDF) for both QT and program execution time in the successfully evaded malware samples.
Figure \ref{fig:query_time} depicts the CDF of the QT, and Figure \ref{fig:query_duration} illustrates the CDF of the program execution time.

In Figure \ref{fig:query_time}, we observe that most successfully evaded malware samples (80\%) against the MaMadroid classifier require only about 6 queries.
Although attacking Drebin and APIGraph necessitates more QT, most successfully evaded malware instances need only about 15 queries.
These results demonstrate that \system is able to perform effectively with a small number of queries.
In Figure \ref{fig:query_duration}, we observe that most successfully evaded malware samples (80\%) against the MaMadroid classifier require under 100 seconds of program execution time.
While attacking Drebin and APIGraph demands more program execution time, most successfully evaded malware instances need only about 500 seconds of program execution time.
To establish the baseline comparison, we apply the MAB and RA against the Drebin, keeping the settings identical to those in the \system.
The results yield 53 and 54 successfully evaded malware instances, respectively.
Subsequently, we delve into a comparative analysis in terms of the average QT and the average program execution time.
Our results reveal that the average QT for \system, MAB, and RA are 9.04, 10.64, and 9.86, respectively.
In terms of the average program execution time, \system, MAB, and RA are clocked at 250.53 seconds, 258.33 seconds, and 305.76 seconds, respectively.
The superior performance of \system can be attributed to the application of the perturbation selection tree and the adjustment policy to some extent.
It is supported by the fact that the sole point of divergence between \system and the baseline methods is their strategy for selecting perturbations.
This finding suggests that \system can efficiently generate adversarial Android malware, indicating high scalability.
Furthermore, \system does not depend on the information shared between malware samples, allowing for parallel implementation, which is another advantage in terms of scalability.


\begin{boxA}
\paragraphbe{Takeaway}
The designs of \system are effective as the attack performance outperforms the baseline methods in terms of the attack success rate.
The QT and the program execution time indicate the low runtime overheads of \system.
\end{boxA}

\subsection{Real-World AVs}
\label{sec:vt}

To assess the performance of \system on real-world AVs, we evaluate it on VirusTotal, which hosts over 70 AVs whose specifics remain undisclosed to us, including large companies such as McAfee, Symantec, and Microsoft.

Specifically, we employ the VirusTotal API~\cite{VirusTotalAPI} to upload applications to VirusTotal and obtain query feedback.
As described in Section~\ref{sec:threatmodel}, we interpret the percentage of detected antivirus engines as malware confidence.
Due to VirusTotal's query limitations and the approximately one-minute analysis time for each uploaded APK, we randomly sample 100 malware samples from our test set for evaluation.
We set 10 query budgets for each malware sample. 
After uploading malware to VirusTotal, we check the analysis results once per minute until the results are available or the duration exceeds five minutes.

\begin{figure}[t]   
	\centering  
	\includegraphics[width=1\linewidth]{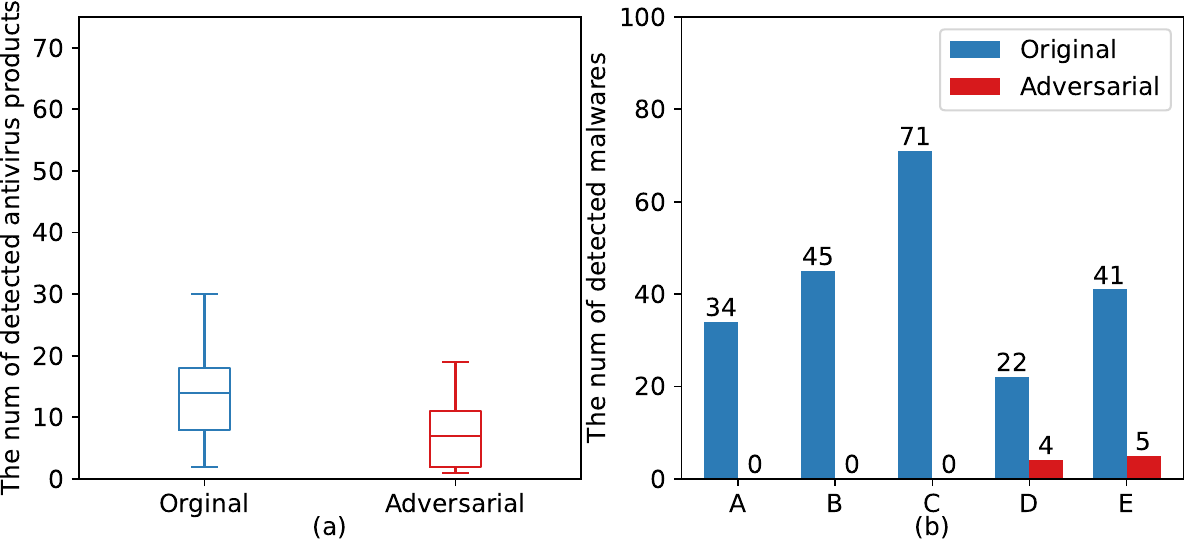}
	\caption{The attack effectiveness of \system against VirusTotal on the primary dataset. (a) The original and adversarial distribution of the number of detected engines. (b) The number of detected malware on 5 antivirus products, where A denotes McAfee, B denotes McAfee-GW-Edition, C denotes SymantecMobileInsight, D denotes Avira, and E denotes Microsoft. The result shows that \system can achieve attack effectiveness against VirusTotal.}
	\label{fig:vt_num}
\end{figure}

In total, we successfully generate 71 adversarial Android malware samples that reduce the number of detected antivirus engines. 
Additionally, 23 malware samples reach the query exception, meaning it has hit the time limit for obtaining the analysis result, and thus, we discontinue further attempts at analyzing it.
Consequently, \system reduces the number of detected antivirus engines for approximately 92\% (71 / (100 - 23)) of the malware samples.
To further analyze attack effectiveness, we compare the original and adversarial distribution of detected engines for the 71 malware samples.
Figure \ref{fig:vt_num} (a) shows that the average detection rate is approximately 18.02\%, corresponding to 13.32 antivirus engines.
In contrast, an average of only 9.84\% (7.28 engines) of antivirus engines flag the \system generated adversarial Android malware as malicious.
Overall, \system decreases the number of detected antivirus engines by 47.98\%, indicating its effectiveness against VirusTotal.

For a more in-depth analysis of \system's attack effectiveness, we examine its performance against five specific antivirus products on VirusTotal provided by the large companies: \textbf{McAfee}, \textbf{McAfee-GW-Edition}, \textbf{SymantecMobileInsight}, \textbf{Avira}, and \textbf{Microsoft}.
We report the number of detected original malware and adversarial malware samples for these five antivirus products.
Figure \ref{fig:vt_num} (b) reveals that \textbf{McAfee}, \textbf{McAfee-GW-Edition}, and \textbf{SymantecMobileInsight} fail to detect any adversarial Android malware generated by \system, indicating a 100\% ASR against these products.
For \textbf{Avira}, \system reduces the detection ratio from 30.98\% to 5.63\%.
For \textbf{Microsoft}, \system decreases the detection ratio from 57.74\% to 7.04\%. 
These results demonstrate that \system can effectively compromise the products of large companies.

\subsubsection{Temporal Effectiveness}
\label{sec:timevalidity}

To assess the effectiveness of \system over time, we carry out evaluations against VirusTotal using the supplementary dataset.
We randomly select 500 malware samples from the supplementary dataset, allocating a query budget of 10 for each malware sample.
The strategy we employ to obtain analysis results for these samples is identical to that used for the malware samples in the primary dataset.

\begin{figure}[t]   
	\centering  
	\includegraphics[width=1\linewidth]{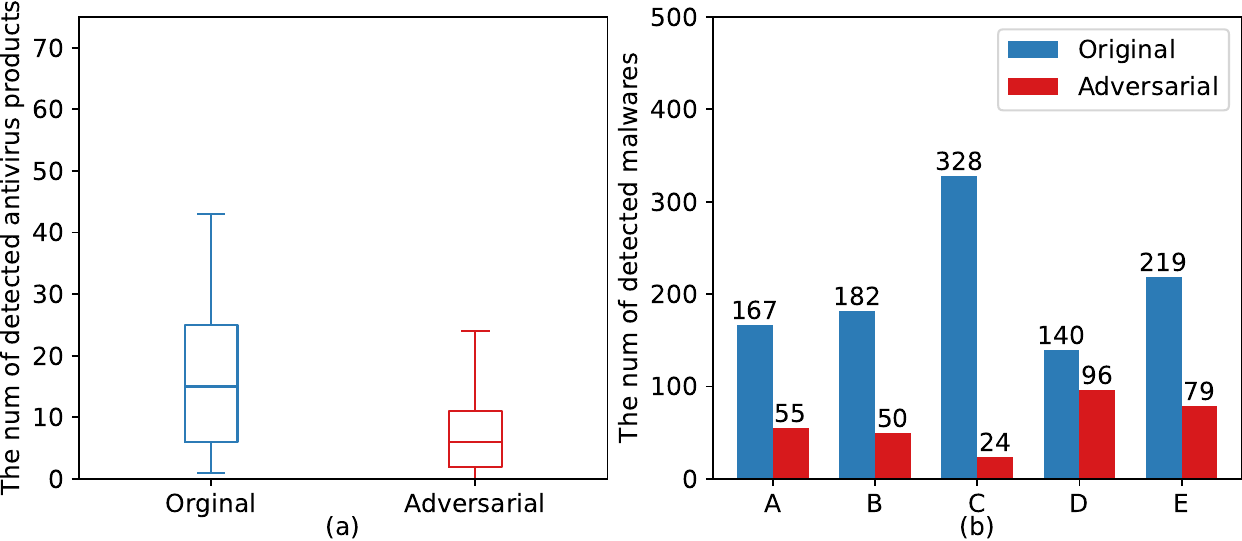}
	\caption{The attack effectiveness of \system against VirusTotal on the supplementary dataset. (a) The original and adversarial distribution of the number of detected engines. (b) The number of detected malware on 5 antivirus products, where A denotes McAfee, B denotes McAfee-GW-Edition, C denotes SymantecMobileInsight, D denotes Avira, and E denotes Microsoft.}
	\label{fig:vt_vs_num}
\end{figure}

In total, we successfully generate 349 adversarial Android malware samples that reduce the number of antivirus engines detecting them.
Furthermore, 109 malware samples reach the query limit, signifying that they have exhausted the time allotment for analysis.
Thus, we terminate further analysis attempts for these samples.
As a result, \system decreases the number of antivirus engines that detect about 89\% (349 / (500 - 109)) of the malware samples.
To delve deeper into the effectiveness of the attack, we compare the original and adversarial distribution of the detected engines for the 349 malware samples.
Figure~\ref{fig:vt_vs_num} (a) indicates that the average detection rate is approximately 22.32\% (16.73 antivirus engines). Conversely, only an average of 11.37\% (8.52 engines) of antivirus engines flag the adversarial Android malware produced by \system as malicious.
In summary, AdvDroidZero reduces the percentage of antivirus engines detecting malware by 56.57\%, underlining its effectiveness over time.

For a more comprehensive analysis of the effectiveness, we assess its performance against the previously mentioned five specific antivirus products.
Figure~\ref{fig:vt_vs_num} (b) proves that \system continues to achieve satisfactory attack effectiveness in most cases against products from large companies.
For example, \system diminishes the detection ratio from 65.60\% to 4.80\% when pitted against \textbf{SymantecMobileInsight}.

\subsection{Dynamic Analysis-based Defense}

In our evaluation of \system, we primarily focus on static analysis-based methods.
This is because a majority (73\% according to a recent survey~\cite{DBLP:journals/csur/LiuTLL23}) of ML-based AMD methods rely on static analysis for feature extraction.
Additionally, static analysis-based methods are likely employed in real-world antivirus systems (Section~\ref{sec:vt}).
The dynamic analysis-based defense is also deployed in the real world.
Therefore, we also evaluate \system against the dynamic analysis-based defense.

To explore the attack effectiveness of \system against the dynamic-based analysis defenses, e.g., sandboxes.
We apply \system to attack the sandboxes in VirusTotal~\cite{VirusTotalSandbox}, which are the SOTA dynamic analysis-based defenses.
Specifically, we randomly select 30 malicious applications that can be detected by the sandboxes in VirusTotal.
We then use \system with a query budget of 10 against VirusTotal to generate the perturbed applications.
Among all the generated perturbed applications, 11 applications are reported by the VirusTotal sandboxes to have no potential malicious behaviors, suggesting that \system achieves an ASR of 37\% against VirusTotal sandboxes within 10 queries.

The results indicate that \system can still have attack effectiveness against dynamic analysis-based defense.
This can be attributed to Android app component perturbations (code perturbations) in the malware perturbation set.
The components in Android applications typically implement independent functions and have rich semantics, making them likely to be triggered by the sandboxes.
Furthermore, since the components are derived from benign applications, they inherently lack malicious behavior.
This may lead the sandbox to spend an excessive amount of time analyzing them, potentially overshadowing the behavior of the original malware.

\subsection{Functionality}

In accordance with prior work~\cite{DBLP:conf/ccs/ZhangZZDCZZY20,DBLP:journals/corr/abs-2303-08509}, we employ both static and dynamic analysis to assess the functionality consistency of \system.

For evaluating functionality consistency through static analysis, we utilize Apktool~\cite{Apktool} to decompile the generated adversarial Android malware.
More explicitly, we randomly sample 20 perturbed malware samples generated in the process against the VirusTotal.
Then, we record every applied perturbation in their generation process and decompile these APKs.
Subsequently, we manually review the decompiled source code of the adversarial APKs.
Our primary objective here is to verify two key points.
First, whether the perturbations have been correctly injected into the malware code.
Second, to establish if the injected code exhibits any relation (i.e., function calls or class relationships) to the original code.
Our results confirm that the decompile process is successful for all perturbed malware samples and that the perturbations are inserted appropriately.
Moreover, we observe no explicit connections between the injected code and the original code.
These observations collectively indicate that the functionality of the perturbed malware remains consistent despite the perturbations.

Regarding dynamic analysis, we install and execute these randomly selected 20 pairs of original and perturbed malware samples from the process against the VirusTotal in an Android Virtual Device (AVD) provided by Android Studio.
The AVD is configured with API level 30 Google APIs and simulates a Nexus 5 device.
Our observations reveal that each malware pair exhibits the same performance and runtime UI.
This outcome suggests that all modified malware samples operate correctly, the inserted functions are not invoked, and thus, the malware's functionality remains unaffected.

\section{Discussion}

\subsection{Possible Defense}

Li \textit{et al}.~\cite{DBLP:conf/www/LiZYLGC21} propose a robust ML-based AMD method called the FD-VAE model, designed to defend against the adversarial Android malware.
The method employs a Variational Autoencoder (VAE) to disentangle features of different classes, thereby enhancing detection performance and robustness.
We re-implement the FD-VAE algorithm using PyTorch~\cite{DBLP:conf/nips/PaszkeGMLBCKLGA19}, following the descriptions and configurations provided in the original paper and their open-source code.
Subsequently, we randomly sample 100 malware samples from the test set in the primary dataset that the FD-VAE can detect for attack evaluation.

The results indicate that \system achieves an ASR of 77\% under 30 query budgets, 50\% under 20 query budgets, and 26\% under 10 query budgets.
These findings demonstrate that \system exhibits strong attack effectiveness when the query budget is relatively large, while the ASR is comparatively lower when the query budget is insufficient.
This can be attributed to the FD-VAE model's use of a rejection mechanism within the VAE model.
The model outputs a malware label when the loss of the VAE surpasses a predefined threshold, which may mislead our algorithm when the query budget is inadequate.

Although \system does not perform optimally in cases with small query budgets, it can still achieve a satisfactory ASR when the query budget is within an acceptable threshold (30 query budgets).

\subsection{Perturbation Types}

To gain insight into the types of perturbations added to our adversarial Android malware, we analyze the distribution of these perturbations in the adversarial Android malware samples we generated.
Specifically, we randomly select 100 malicious applications that can be detected by VirusTotal.
We then apply \system to these applications, attacking VirusTotal with a query budget of 10 for every sample.
Consequently, we successfully generate 61 adversarial Android malware samples that manage to reduce the number of antivirus engines that detect them on VirusTotal.

\begin{figure}[t]   
	\centering  
	\includegraphics[width=1\linewidth]{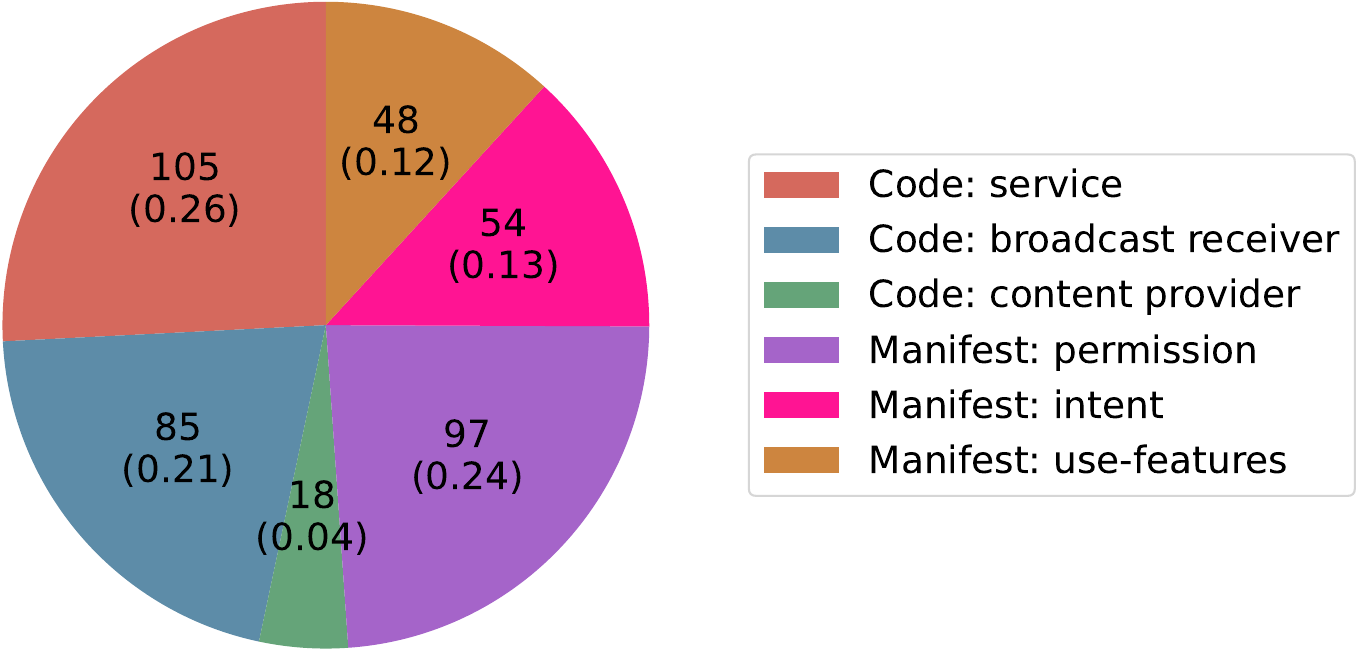}
	\caption{The distribution of the added perturbation type in our generated adversarial Android malware.}
	\label{fig:PerturbDis}
\end{figure}

As illustrated in Figure~\ref{fig:PerturbDis}, the most common perturbations are primarily service, permission, and broadcast receivers, with respective occurrence ratios of 26\%, 24\%, and 21\%.
This can potentially be attributed to the fact that services and broadcast receivers are widely utilized in benign applications (as depicted in Figure~\ref{fig:num_components}), causing these features to be identified by ML-based AMD methods.
As a result, they aid malicious applications in evading detection.
Permissions have been extensively studied and identified as a critical feature in malware detection~\cite{DBLP:conf/ccs/AuZHL12}.
In designing \system, we consciously decide to avoid integrating permissions that are considered high-risk, opting instead to focus on the permissions that have normal or signature protection levels.
This approach confers a seemingly benign appearance upon the perturbed malicious applications, assisting them in avoiding detection.

\subsection{Limitations \& Future Work}
\label{sec:limitation}

In this study, we introduce an efficient query-based attack framework, \system, targeting ML-based AMD methods under zero knowledge setting. 
Our work aims to offer valuable insights into the field of AMD and to raise awareness regarding the potential threats posed by attacks.
Furthermore, our method can be employed to assess the robustness of existing ML-based AMD methods under zero knowledge setting.
In the following, we discuss some limitations of our approach and outline potential future research directions.

\paragraphbe{Obscured Model Output}
The \system framework needs the confidence of the target model provided by ML-based AMD methods~\cite{DBLP:conf/ndss/ArpSHGR14,DBLP:journals/tosem/WuCGFLWL21}.
However, in some cases, the model output might be restricted or obscured, which can limit the effectiveness of \system.
To adapt \system to these limited output scenarios, some strategies can be implemented, such as approximating the confidence level by the percentage of detected antivirus engines provided in VirusTotal.
This technique can offer a reasonable estimate when the confidence of the ML-based AMD method is unavailable.
Nevertheless, we acknowledge that there might be certain ML-based AMD methods that are not susceptible to the \system framework due to the constraints in inspecting output.
In future research, we plan to tackle these specific cases.

\paragraphbe{Hybrid Analysis-based Defense}
Our primary goal is to design an effective and efficient attack framework for generating adversarial Android malware in a practical setting.
ML-based AMD methods that employ hybrid program analysis are a promising avenue for research.
These methods have the potential to combine the strengths of both static and dynamic program analysis.
However, according to a recent survey~\cite{DBLP:journals/csur/LiuTLL23}, these methods have yet to be thoroughly explored and have many unresolved challenges, such as high computational demands, etc.
Thus, the application of them remains limited.
We plan to explore the impact of hybrid analysis-based defenses in future research.

\paragraphbe{Malware Perturbation Set Limitation}
The \system framework relies on the malware perturbation set.
If none of the perturbations affect the feature values of the target AMD methods, the efficacy of \system would be compromised.
However, we contend that such a scenario is unlikely to occur in real-world settings for several reasons.
Firstly, our perturbation set is relatively general and has demonstrated its effectiveness in influencing real-world antivirus solutions.
Moreover, extending the perturbation selection tree is straightforward, as it merely requires the addition of a subtree to accommodate new perturbations.
This inherent flexibility and adaptability of \system make it a robust framework to attack against the ML-based AMD methods.

\subsection{Potential Ethical Concerns}
\label{sec:ethic}
The primary objective of our study is to assess the robustness of ML-based AMD methods, a topic with established precedence in earlier works~\cite{DBLP:conf/sp/PierazziPCC20,DBLP:conf/ccs/ZhaoZZZZLYYL21,DBLP:journals/corr/abs-2303-08509}.
This is driven by the potential for adversaries to engineer malware applications that can evade detection, thus underscoring the necessity for robust ML-based AMD methods.
Even though the intent is strict about evaluating the robustness of ML-based AMD methods, potential ethical concerns are associated with our research.
Therefore, we will limit code sharing upon request to verified academic researchers only, following a precedent established by previous studies~\cite{DBLP:conf/sp/PierazziPCC20,DBLP:conf/ccs/ZhaoZZZZLYYL21}.
Besides, in VirusTotal experiments, we utilized VirusTotal in the same capacity as an ordinary user would—submitting applications.
It aids the security community by providing samples of this type of malicious application.

\section{Related Work}
\label{sec:relatedwork}

Adversarial examples have recently garnered significant attention in various domains, such as text classification \cite{DBLP:conf/ndss/LiJDLW19,DBLP:conf/ccs/DuJSZLSFYB021}, reinforcement learning \cite{DBLP:conf/uss/Wu0WX21}, and explainability \cite{DBLP:conf/uss/ZhangWSJL020}. 
In the malware domain, numerous attack algorithms \cite{DBLP:conf/sp/PierazziPCC20,DBLP:conf/ccs/ZhaoZZZZLYYL21,DBLP:journals/corr/abs-2303-08509,DBLP:journals/tifs/0002LWW0N0020,DBLP:journals/tifs/LiL20,DBLP:journals/corr/abs-2110-03301,DBLP:conf/asiaccs/SongLAGKY22,DBLP:journals/corr/abs-2111-10085} have been developed to attack malware detection methods.
For instance, Sun \textit{et al}. \cite{DBLP:journals/corr/abs-2110-03301} employ machine learning explainability techniques to select benign features for modification, while Song \textit{et al}. \cite{DBLP:conf/asiaccs/SongLAGKY22} approach the attack problem as a multi-armed bandit problem, creating a generic machine and a specific machine to profile perturbations using Thompson sampling to cope with the delayed feedback.

In the Android malware domain, Chen \textit{et al}. \cite{DBLP:journals/tifs/0002LWW0N0020} utilize optimization algorithms to generate adversarial perturbations in the feature space and introduce a method for applying optimal perturbations to APKs.
Pierazzi \textit{et al}. \cite{DBLP:conf/sp/PierazziPCC20} propose the APG (adversarial problem generation) framework, which generates adversarial malware from feature space to problem space.
Zhao \textit{et al}. \cite{DBLP:conf/ccs/ZhaoZZZZLYYL21} present a structural attack that employs reinforcement learning to target FCG-based AMD methods.
Li \textit{et al}. \cite{DBLP:journals/tifs/LiL20} explore ensemble learning algorithms for adversarial Android malware.
Recently, Li \textit{et al}. \cite{DBLP:journals/corr/abs-2303-08509} developed the BagAmmo algorithm to create adversarial Android malware against FCG-based AMD methods in the limited knowledge setting.
Bostani \textit{et al}. \cite{DBLP:journals/corr/abs-2110-03301} introduce EvadeDroid, which uses a random search algorithm to inject instruction-level gadgets into malware.

However, these methods do not fully address the practical requirements for generating adversarial Android malware.
For instance, they may necessitate some knowledge (feature space, model parameters, training dataset) about the target AMD method \cite{DBLP:conf/sp/PierazziPCC20,DBLP:journals/tifs/0002LWW0N0020,DBLP:journals/tifs/LiL20,DBLP:conf/ccs/ZhaoZZZZLYYL21}, or be limited to specific AMD models \cite{DBLP:conf/ccs/ZhaoZZZZLYYL21,DBLP:journals/corr/abs-2303-08509}.
Additionally, they may require a large number of queries for the target model \cite{DBLP:journals/corr/abs-2303-08509}, and/or their modifications in the problem space may not be robust to pre-processing \cite{DBLP:journals/corr/abs-2110-03301}.

\section{Conclusion}

This paper introduces \system, an efficient query-based attack framework designed to generate adversarial Android malware under zero knowledge setting. 
To address the vast and heterogeneous search space, \system employs an innovative data structure termed as the perturbation selection tree and proposes an adjustment policy for efficiently choosing the appropriate perturbations.
Our experimental results, conducted on the large-scale real-world datasets, demonstrate that \system is effective against a wide variety of ML-based AMD methods using static program analysis.
Furthermore, we evaluate \system's performance against real-world antivirus solutions, sandboxes and robust AMD methods, finding that it can bypass these defenses.
The findings underscore the potential of \system as a valuable framework in the field of attacks on AMD methods.

\begin{acks}
We sincerely appreciate our shepherd and the anonymous reviewers for their insightful comments.
We would like to extend our gratitude to Chenghui Shi, Changjiang Li, Qinying Wang, and Yuhao Mao for their thoughtful feedback.
We thank the support from the Zhejiang University College of Computer Science and Technology and the Zhejiang University NGICS Platform.
This work was partly supported by NSFC under No. U1936215 and the Fundamental Research Funds for the Central Universities (Zhejiang University NGICS Platform).
\end{acks}


\bibliographystyle{ACM-Reference-Format}
\bibliography{reference}


\begin{thebibliography}{81}


\ifx \showCODEN    \undefined \def \showCODEN     #1{\unskip}     \fi
\ifx \showDOI      \undefined \def \showDOI       #1{#1}\fi
\ifx \showISBNx    \undefined \def \showISBNx     #1{\unskip}     \fi
\ifx \showISBNxiii \undefined \def \showISBNxiii  #1{\unskip}     \fi
\ifx \showISSN     \undefined \def \showISSN      #1{\unskip}     \fi
\ifx \showLCCN     \undefined \def \showLCCN      #1{\unskip}     \fi
\ifx \shownote     \undefined \def \shownote      #1{#1}          \fi
\ifx \showarticletitle \undefined \def \showarticletitle #1{#1}   \fi
\ifx \showURL      \undefined \def \showURL       {\relax}        \fi
\providecommand\bibfield[2]{#2}
\providecommand\bibinfo[2]{#2}
\providecommand\natexlab[1]{#1}
\providecommand\showeprint[2][]{arXiv:#2}

\bibitem[\protect\citeauthoryear{Allix, Bissyand{\'{e}}, Klein, and
  Traon}{Allix et~al\mbox{.}}{2016}]%
        {DBLP:conf/msr/AllixBKT16}
\bibfield{author}{\bibinfo{person}{Kevin Allix},
  \bibinfo{person}{Tegawend{\'{e}}~F. Bissyand{\'{e}}},
  \bibinfo{person}{Jacques Klein}, {and} \bibinfo{person}{Yves~Le Traon}.}
  \bibinfo{year}{2016}\natexlab{}.
\newblock \showarticletitle{AndroZoo: collecting millions of Android apps for
  the research community}. In \bibinfo{booktitle}{\emph{{MSR}}}.
\newblock


\bibitem[\protect\citeauthoryear{Androguard}{Androguard}{2023}]%
        {Androguard}
\bibfield{author}{\bibinfo{person}{Androguard}.}
  \bibinfo{year}{2023}\natexlab{}.
\newblock \bibinfo{title}{Androguard}.
\newblock
  \bibinfo{howpublished}{\url{https://github.com/androguard/androguard}}.
\newblock
\newblock
\shownote{[Accessed on Apr. 21, 2023].}


\bibitem[\protect\citeauthoryear{Android}{Android}{2023a}]%
        {AndroidDocumentation}
\bibfield{author}{\bibinfo{person}{Android}.} \bibinfo{year}{2023}\natexlab{a}.
\newblock \bibinfo{title}{{Android Documentation}}.
\newblock
\newblock
\newblock
\shownote{[Accessed on Apr. 13, 2023].}


\bibitem[\protect\citeauthoryear{Android}{Android}{2023b}]%
        {AndoridManifestDocumentation}
\bibfield{author}{\bibinfo{person}{Android}.} \bibinfo{year}{2023}\natexlab{b}.
\newblock \bibinfo{title}{Android Manifest Documentation}.
\newblock
  \bibinfo{howpublished}{\url{https://developer.android.com/guide/topics/manifest/manifest-intro}}.
\newblock
\newblock
\shownote{[Accessed on Apr. 21, 2023].}


\bibitem[\protect\citeauthoryear{Android}{Android}{2023c}]%
        {AndoridPermission}
\bibfield{author}{\bibinfo{person}{Android}.} \bibinfo{year}{2023}\natexlab{c}.
\newblock \bibinfo{title}{Android Permission}.
\newblock
  \bibinfo{howpublished}{\url{https://developer.android.com/reference/android/Manifest.permission}}.
\newblock
\newblock
\shownote{[Accessed on Apr. 21, 2023].}


\bibitem[\protect\citeauthoryear{Android}{Android}{2023d}]%
        {AppComponents}
\bibfield{author}{\bibinfo{person}{Android}.} \bibinfo{year}{2023}\natexlab{d}.
\newblock \bibinfo{title}{App Components}.
\newblock
  \bibinfo{howpublished}{\url{https://developer.android.com/guide/components/fundamentals}}.
\newblock
\newblock
\shownote{[Accessed on Apr. 21, 2023].}


\bibitem[\protect\citeauthoryear{Android}{Android}{2023e}]%
        {AndroidIntent}
\bibfield{author}{\bibinfo{person}{Android}.} \bibinfo{year}{2023}\natexlab{e}.
\newblock \bibinfo{title}{App Intent}.
\newblock
  \bibinfo{howpublished}{\url{https://developer.android.com/reference/android/content/Intent}}.
\newblock
\newblock
\shownote{[Accessed on Apr. 21, 2023].}


\bibitem[\protect\citeauthoryear{Android}{Android}{2023f}]%
        {AndroidProcess}
\bibfield{author}{\bibinfo{person}{Android}.} \bibinfo{year}{2023}\natexlab{f}.
\newblock \bibinfo{title}{App Process}.
\newblock
  \bibinfo{howpublished}{\url{https://developer.android.com/guide/components/processes-and-threads}}.
\newblock
\newblock
\shownote{[Accessed on Apr. 21, 2023].}


\bibitem[\protect\citeauthoryear{Android}{Android}{2023g}]%
        {ReferencedFeatures}
\bibfield{author}{\bibinfo{person}{Android}.} \bibinfo{year}{2023}\natexlab{g}.
\newblock \bibinfo{title}{Use-Feature reference}.
\newblock
  \bibinfo{howpublished}{\url{https://developer.android.com/guide/topics/manifest/uses-feature-element}}.
\newblock
\newblock
\shownote{[Accessed on Apr. 21, 2023].}


\bibitem[\protect\citeauthoryear{Apktool}{Apktool}{2023}]%
        {Apktool}
\bibfield{author}{\bibinfo{person}{Apktool}.} \bibinfo{year}{2023}\natexlab{}.
\newblock \bibinfo{title}{Apktool}.
\newblock \bibinfo{howpublished}{\url{https://ibotpeaches.github.io/Apktool/}}.
\newblock
\newblock
\shownote{[Accessed on Apr. 21, 2023].}


\bibitem[\protect\citeauthoryear{Apruzzese, Anderson, Dambra, Freeman,
  Pierazzi, and Roundy}{Apruzzese et~al\mbox{.}}{2022}]%
        {DBLP:journals/corr/abs-2212-14315}
\bibfield{author}{\bibinfo{person}{Giovanni Apruzzese},
  \bibinfo{person}{Hyrum~S. Anderson}, \bibinfo{person}{Savino Dambra},
  \bibinfo{person}{David Freeman}, \bibinfo{person}{Fabio Pierazzi}, {and}
  \bibinfo{person}{Kevin~A. Roundy}.} \bibinfo{year}{2022}\natexlab{}.
\newblock \showarticletitle{"Real Attackers Don't Compute Gradients": Bridging
  the Gap Between Adversarial {ML} Research and Practice}.
\newblock \bibinfo{journal}{\emph{CoRR}} (\bibinfo{year}{2022}).
\newblock


\bibitem[\protect\citeauthoryear{Arp, Quiring, Pendlebury, Warnecke, Pierazzi,
  Wressnegger, Cavallaro, and Rieck}{Arp et~al\mbox{.}}{2022}]%
        {DBLP:conf/uss/ArpQPWPWCR22}
\bibfield{author}{\bibinfo{person}{Daniel Arp}, \bibinfo{person}{Erwin
  Quiring}, \bibinfo{person}{Feargus Pendlebury}, \bibinfo{person}{Alexander
  Warnecke}, \bibinfo{person}{Fabio Pierazzi}, \bibinfo{person}{Christian
  Wressnegger}, \bibinfo{person}{Lorenzo Cavallaro}, {and}
  \bibinfo{person}{Konrad Rieck}.} \bibinfo{year}{2022}\natexlab{}.
\newblock \showarticletitle{Dos and Don'ts of Machine Learning in Computer
  Security}. In \bibinfo{booktitle}{\emph{{USENIX} Security Symposium}}.
\newblock


\bibitem[\protect\citeauthoryear{Arp, Spreitzenbarth, Hubner, Gascon, and
  Rieck}{Arp et~al\mbox{.}}{2014}]%
        {DBLP:conf/ndss/ArpSHGR14}
\bibfield{author}{\bibinfo{person}{Daniel Arp}, \bibinfo{person}{Michael
  Spreitzenbarth}, \bibinfo{person}{Malte Hubner}, \bibinfo{person}{Hugo
  Gascon}, {and} \bibinfo{person}{Konrad Rieck}.}
  \bibinfo{year}{2014}\natexlab{}.
\newblock \showarticletitle{{DREBIN:} Effective and Explainable Detection of
  Android Malware in Your Pocket}. In \bibinfo{booktitle}{\emph{{NDSS}}}.
\newblock


\bibitem[\protect\citeauthoryear{Arzt, Rasthofer, Fritz, Bodden, Bartel, Klein,
  Traon, Octeau, and McDaniel}{Arzt et~al\mbox{.}}{2014}]%
        {DBLP:conf/pldi/ArztRFBBKTOM14}
\bibfield{author}{\bibinfo{person}{Steven Arzt}, \bibinfo{person}{Siegfried
  Rasthofer}, \bibinfo{person}{Christian Fritz}, \bibinfo{person}{Eric Bodden},
  \bibinfo{person}{Alexandre Bartel}, \bibinfo{person}{Jacques Klein},
  \bibinfo{person}{Yves~Le Traon}, \bibinfo{person}{Damien Octeau}, {and}
  \bibinfo{person}{Patrick~D. McDaniel}.} \bibinfo{year}{2014}\natexlab{}.
\newblock \showarticletitle{FlowDroid: precise context, flow, field,
  object-sensitive and lifecycle-aware taint analysis for Android apps}. In
  \bibinfo{booktitle}{\emph{{PLDI}}}.
\newblock


\bibitem[\protect\citeauthoryear{Au, Zhou, Huang, and Lie}{Au
  et~al\mbox{.}}{2012}]%
        {DBLP:conf/ccs/AuZHL12}
\bibfield{author}{\bibinfo{person}{Kathy Wain~Yee Au}, \bibinfo{person}{Yi~Fan
  Zhou}, \bibinfo{person}{Zhen Huang}, {and} \bibinfo{person}{David Lie}.}
  \bibinfo{year}{2012}\natexlab{}.
\newblock \showarticletitle{PScout: analyzing the Android permission
  specification}. In \bibinfo{booktitle}{\emph{ACM {CCS}}}.
\newblock


\bibitem[\protect\citeauthoryear{AV-ATLAS}{AV-ATLAS}{2023}]%
        {MalwareSample}
\bibfield{author}{\bibinfo{person}{AV-ATLAS}.} \bibinfo{year}{2023}\natexlab{}.
\newblock \bibinfo{title}{{Total Amount of Android Malware}}.
\newblock
  \bibinfo{howpublished}{\url{https://portal.av-atlas.org/malware/statistics}}.
\newblock
\newblock
\shownote{[Accessed on Apr. 13, 2023].}


\bibitem[\protect\citeauthoryear{Barbero, Pendlebury, Pierazzi, and
  Cavallaro}{Barbero et~al\mbox{.}}{2022}]%
        {DBLP:conf/sp/BarberoPPC22}
\bibfield{author}{\bibinfo{person}{Federico Barbero}, \bibinfo{person}{Feargus
  Pendlebury}, \bibinfo{person}{Fabio Pierazzi}, {and} \bibinfo{person}{Lorenzo
  Cavallaro}.} \bibinfo{year}{2022}\natexlab{}.
\newblock \showarticletitle{Transcending {TRANSCEND:} Revisiting Malware
  Classification in the Presence of Concept Drift}. In
  \bibinfo{booktitle}{\emph{{IEEE} S\&P}}.
\newblock


\bibitem[\protect\citeauthoryear{Biggio and Roli}{Biggio and Roli}{2018}]%
        {DBLP:journals/pr/BiggioR18}
\bibfield{author}{\bibinfo{person}{Battista Biggio} {and}
  \bibinfo{person}{Fabio Roli}.} \bibinfo{year}{2018}\natexlab{}.
\newblock \showarticletitle{Wild patterns: Ten years after the rise of
  adversarial machine learning}.
\newblock \bibinfo{journal}{\emph{Pattern Recognit.}} (\bibinfo{year}{2018}).
\newblock


\bibitem[\protect\citeauthoryear{Bostani and Moonsamy}{Bostani and
  Moonsamy}{2021}]%
        {DBLP:journals/corr/abs-2110-03301}
\bibfield{author}{\bibinfo{person}{Hamid Bostani} {and}
  \bibinfo{person}{Veelasha Moonsamy}.} \bibinfo{year}{2021}\natexlab{}.
\newblock \showarticletitle{EvadeDroid: {A} Practical Evasion Attack on Machine
  Learning for Black-box Android Malware Detection}.
\newblock \bibinfo{journal}{\emph{CoRR}} (\bibinfo{year}{2021}).
\newblock


\bibitem[\protect\citeauthoryear{Brendel, Rauber, and Bethge}{Brendel
  et~al\mbox{.}}{2018}]%
        {DBLP:conf/iclr/BrendelRB18}
\bibfield{author}{\bibinfo{person}{Wieland Brendel}, \bibinfo{person}{Jonas
  Rauber}, {and} \bibinfo{person}{Matthias Bethge}.}
  \bibinfo{year}{2018}\natexlab{}.
\newblock \showarticletitle{Decision-Based Adversarial Attacks: Reliable
  Attacks Against Black-Box Machine Learning Models}. In
  \bibinfo{booktitle}{\emph{ICLR}}.
\newblock


\bibitem[\protect\citeauthoryear{Carlini, Athalye, Papernot, Brendel, Rauber,
  Tsipras, Goodfellow, Madry, and Kurakin}{Carlini et~al\mbox{.}}{2019}]%
        {DBLP:journals/corr/abs-1902-06705}
\bibfield{author}{\bibinfo{person}{Nicholas Carlini}, \bibinfo{person}{Anish
  Athalye}, \bibinfo{person}{Nicolas Papernot}, \bibinfo{person}{Wieland
  Brendel}, \bibinfo{person}{Jonas Rauber}, \bibinfo{person}{Dimitris Tsipras},
  \bibinfo{person}{Ian~J. Goodfellow}, \bibinfo{person}{Aleksander Madry},
  {and} \bibinfo{person}{Alexey Kurakin}.} \bibinfo{year}{2019}\natexlab{}.
\newblock \showarticletitle{On Evaluating Adversarial Robustness}.
\newblock \bibinfo{journal}{\emph{CoRR}} (\bibinfo{year}{2019}).
\newblock


\bibitem[\protect\citeauthoryear{Carlini and Wagner}{Carlini and
  Wagner}{2017}]%
        {DBLP:conf/sp/Carlini017}
\bibfield{author}{\bibinfo{person}{Nicholas Carlini} {and}
  \bibinfo{person}{David~A. Wagner}.} \bibinfo{year}{2017}\natexlab{}.
\newblock \showarticletitle{Towards Evaluating the Robustness of Neural
  Networks}. In \bibinfo{booktitle}{\emph{{IEEE} S\&P}}.
\newblock


\bibitem[\protect\citeauthoryear{Chatterjee, Doerfler, Orgad, Havron, Palmer,
  Freed, Levy, Dell, McCoy, and Ristenpart}{Chatterjee et~al\mbox{.}}{2018}]%
        {DBLP:conf/sp/ChatterjeeDOHPF18}
\bibfield{author}{\bibinfo{person}{Rahul Chatterjee},
  \bibinfo{person}{Periwinkle Doerfler}, \bibinfo{person}{Hadas Orgad},
  \bibinfo{person}{Sam Havron}, \bibinfo{person}{Jackeline Palmer},
  \bibinfo{person}{Diana Freed}, \bibinfo{person}{Karen Levy},
  \bibinfo{person}{Nicola Dell}, \bibinfo{person}{Damon McCoy}, {and}
  \bibinfo{person}{Thomas Ristenpart}.} \bibinfo{year}{2018}\natexlab{}.
\newblock \showarticletitle{The Spyware Used in Intimate Partner Violence}. In
  \bibinfo{booktitle}{\emph{{IEEE} Symposium on Security and Privacy}}.
\newblock


\bibitem[\protect\citeauthoryear{Chen, Jordan, and Wainwright}{Chen
  et~al\mbox{.}}{2020a}]%
        {DBLP:conf/sp/ChenJW20}
\bibfield{author}{\bibinfo{person}{Jianbo Chen}, \bibinfo{person}{Michael~I.
  Jordan}, {and} \bibinfo{person}{Martin~J. Wainwright}.}
  \bibinfo{year}{2020}\natexlab{a}.
\newblock \showarticletitle{HopSkipJumpAttack: {A} Query-Efficient
  Decision-Based Attack}. In \bibinfo{booktitle}{\emph{{IEEE} S\&P}}.
\newblock


\bibitem[\protect\citeauthoryear{Chen, Li, Wang, Wen, Zhang, Nepal, Xiang, and
  Ren}{Chen et~al\mbox{.}}{2020b}]%
        {DBLP:journals/tifs/0002LWW0N0020}
\bibfield{author}{\bibinfo{person}{Xiao Chen}, \bibinfo{person}{Chaoran Li},
  \bibinfo{person}{Derui Wang}, \bibinfo{person}{Sheng Wen},
  \bibinfo{person}{Jun Zhang}, \bibinfo{person}{Surya Nepal},
  \bibinfo{person}{Yang Xiang}, {and} \bibinfo{person}{Kui Ren}.}
  \bibinfo{year}{2020}\natexlab{b}.
\newblock \showarticletitle{Android {HIV:} {A} Study of Repackaging Malware for
  Evading Machine-Learning Detection}.
\newblock \bibinfo{journal}{\emph{{IEEE} Trans. Inf. Forensics Secur.}}
  (\bibinfo{year}{2020}).
\newblock


\bibitem[\protect\citeauthoryear{Daoudi, Allix, Bissyand{\'{e}}, and
  Klein}{Daoudi et~al\mbox{.}}{2022}]%
        {DBLP:journals/tissec/DaoudiABK22}
\bibfield{author}{\bibinfo{person}{Nadia Daoudi}, \bibinfo{person}{Kevin
  Allix}, \bibinfo{person}{Tegawend{\'{e}} Fran{\c{c}}ois D.~Assise
  Bissyand{\'{e}}}, {and} \bibinfo{person}{Jacques Klein}.}
  \bibinfo{year}{2022}\natexlab{}.
\newblock \showarticletitle{A Deep Dive Inside {DREBIN:} An Explorative
  Analysis beyond Android Malware Detection Scores}.
\newblock \bibinfo{journal}{\emph{{ACM} Trans. Priv. Secur.}}
  (\bibinfo{year}{2022}).
\newblock


\bibitem[\protect\citeauthoryear{Demontis, Melis, Biggio, Maiorca, Arp, Rieck,
  Corona, Giacinto, and Roli}{Demontis et~al\mbox{.}}{2019}]%
        {DBLP:journals/tdsc/DemontisMBMARCG19}
\bibfield{author}{\bibinfo{person}{Ambra Demontis}, \bibinfo{person}{Marco
  Melis}, \bibinfo{person}{Battista Biggio}, \bibinfo{person}{Davide Maiorca},
  \bibinfo{person}{Daniel Arp}, \bibinfo{person}{Konrad Rieck},
  \bibinfo{person}{Igino Corona}, \bibinfo{person}{Giorgio Giacinto}, {and}
  \bibinfo{person}{Fabio Roli}.} \bibinfo{year}{2019}\natexlab{}.
\newblock \showarticletitle{Yes, Machine Learning Can Be More Secure! {A} Case
  Study on Android Malware Detection}.
\newblock \bibinfo{journal}{\emph{{IEEE} Trans. Dependable Secur. Comput.}}
  (\bibinfo{year}{2019}).
\newblock


\bibitem[\protect\citeauthoryear{Du, Ji, Shen, Zhang, Li, Shi, Fang, Yin,
  Beyah, and Wang}{Du et~al\mbox{.}}{2021}]%
        {DBLP:conf/ccs/DuJSZLSFYB021}
\bibfield{author}{\bibinfo{person}{Tianyu Du}, \bibinfo{person}{Shouling Ji},
  \bibinfo{person}{Lujia Shen}, \bibinfo{person}{Yao Zhang},
  \bibinfo{person}{Jinfeng Li}, \bibinfo{person}{Jie Shi},
  \bibinfo{person}{Chengfang Fang}, \bibinfo{person}{Jianwei Yin},
  \bibinfo{person}{Raheem Beyah}, {and} \bibinfo{person}{Ting Wang}.}
  \bibinfo{year}{2021}\natexlab{}.
\newblock \showarticletitle{Cert-RNN: Towards Certifying the Robustness of
  Recurrent Neural Networks}. In \bibinfo{booktitle}{\emph{ACM {CCS}}}.
\newblock


\bibitem[\protect\citeauthoryear{Faruki, Bharmal, Laxmi, Ganmoor, Gaur, Conti,
  and Rajarajan}{Faruki et~al\mbox{.}}{2015}]%
        {DBLP:journals/comsur/FarukiBLGGCR15}
\bibfield{author}{\bibinfo{person}{Parvez Faruki}, \bibinfo{person}{Ammar
  Bharmal}, \bibinfo{person}{Vijay Laxmi}, \bibinfo{person}{Vijay Ganmoor},
  \bibinfo{person}{Manoj~Singh Gaur}, \bibinfo{person}{Mauro Conti}, {and}
  \bibinfo{person}{Muttukrishnan Rajarajan}.} \bibinfo{year}{2015}\natexlab{}.
\newblock \showarticletitle{Android Security: {A} Survey of Issues, Malware
  Penetration, and Defenses}.
\newblock \bibinfo{journal}{\emph{{IEEE} Commun. Surv. Tutorials}}
  (\bibinfo{year}{2015}).
\newblock


\bibitem[\protect\citeauthoryear{Fu, Dong, Su, Zhu, and Zhang}{Fu
  et~al\mbox{.}}{2022}]%
        {DBLP:conf/uss/FuD00022}
\bibfield{author}{\bibinfo{person}{Qi{-}An Fu}, \bibinfo{person}{Yinpeng Dong},
  \bibinfo{person}{Hang Su}, \bibinfo{person}{Jun Zhu}, {and}
  \bibinfo{person}{Chao Zhang}.} \bibinfo{year}{2022}\natexlab{}.
\newblock \showarticletitle{AutoDA: Automated Decision-based Iterative
  Adversarial Attacks}. In \bibinfo{booktitle}{\emph{{USENIX} Security
  Symposium}}.
\newblock


\bibitem[\protect\citeauthoryear{{Google Play}}{{Google Play}}{2023}]%
        {GooglePlay}
\bibfield{author}{\bibinfo{person}{{Google Play}}.}
  \bibinfo{year}{2023}\natexlab{}.
\newblock \bibinfo{title}{{Google Play}}.
\newblock
\newblock
\newblock
\shownote{[Accessed on Apr. 13, 2023].}


\bibitem[\protect\citeauthoryear{Grosse, Papernot, Manoharan, Backes, and
  McDaniel}{Grosse et~al\mbox{.}}{2017}]%
        {DBLP:conf/esorics/GrossePMBM17}
\bibfield{author}{\bibinfo{person}{Kathrin Grosse}, \bibinfo{person}{Nicolas
  Papernot}, \bibinfo{person}{Praveen Manoharan}, \bibinfo{person}{Michael
  Backes}, {and} \bibinfo{person}{Patrick~D. McDaniel}.}
  \bibinfo{year}{2017}\natexlab{}.
\newblock \showarticletitle{Adversarial Examples for Malware Detection}. In
  \bibinfo{booktitle}{\emph{{ESORICS}}}.
\newblock


\bibitem[\protect\citeauthoryear{Guo, Mu, Xu, Su, Wang, and Xing}{Guo
  et~al\mbox{.}}{2018}]%
        {DBLP:conf/ccs/GuoMXSWX18}
\bibfield{author}{\bibinfo{person}{Wenbo Guo}, \bibinfo{person}{Dongliang Mu},
  \bibinfo{person}{Jun Xu}, \bibinfo{person}{Purui Su}, \bibinfo{person}{Gang
  Wang}, {and} \bibinfo{person}{Xinyu Xing}.} \bibinfo{year}{2018}\natexlab{}.
\newblock \showarticletitle{{LEMNA:} Explaining Deep Learning based Security
  Applications}. In \bibinfo{booktitle}{\emph{ACM {CCS}}}.
\newblock


\bibitem[\protect\citeauthoryear{Jordaney, Sharad, Dash, Wang, Papini,
  Nouretdinov, and Cavallaro}{Jordaney et~al\mbox{.}}{2017}]%
        {DBLP:conf/uss/JordaneySDWPNC17}
\bibfield{author}{\bibinfo{person}{Roberto Jordaney}, \bibinfo{person}{Kumar
  Sharad}, \bibinfo{person}{Santanu~Kumar Dash}, \bibinfo{person}{Zhi Wang},
  \bibinfo{person}{Davide Papini}, \bibinfo{person}{Ilia Nouretdinov}, {and}
  \bibinfo{person}{Lorenzo Cavallaro}.} \bibinfo{year}{2017}\natexlab{}.
\newblock \showarticletitle{Transcend: Detecting Concept Drift in Malware
  Classification Models}. In \bibinfo{booktitle}{\emph{{USENIX} Security
  Symposium}}.
\newblock


\bibitem[\protect\citeauthoryear{Karer and Soni}{Karer and Soni}{2015}]%
        {7475289}
\bibfield{author}{\bibinfo{person}{Hiral~H. Karer} {and}
  \bibinfo{person}{Purvi~B. Soni}.} \bibinfo{year}{2015}\natexlab{}.
\newblock \showarticletitle{Dead code elimination technique in eclipse compiler
  for Java}. In \bibinfo{booktitle}{\emph{ICCICCT}}.
\newblock


\bibitem[\protect\citeauthoryear{Li, Ji, Weng, Li, Shi, Beyah, Guo, Wang, and
  Wang}{Li et~al\mbox{.}}{2022a}]%
        {DBLP:journals/tdsc/LiJWLSBGWW22}
\bibfield{author}{\bibinfo{person}{Changjiang Li}, \bibinfo{person}{Shouling
  Ji}, \bibinfo{person}{Haiqin Weng}, \bibinfo{person}{Bo Li},
  \bibinfo{person}{Jie Shi}, \bibinfo{person}{Raheem Beyah},
  \bibinfo{person}{Shanqing Guo}, \bibinfo{person}{Zonghui Wang}, {and}
  \bibinfo{person}{Ting Wang}.} \bibinfo{year}{2022}\natexlab{a}.
\newblock \showarticletitle{Towards Certifying the Asymmetric Robustness for
  Neural Networks: Quantification and Applications}.
\newblock \bibinfo{journal}{\emph{{IEEE} Trans. Dependable Secur. Comput.}}
  (\bibinfo{year}{2022}).
\newblock


\bibitem[\protect\citeauthoryear{Li and Li}{Li and Li}{2020}]%
        {DBLP:journals/tifs/LiL20}
\bibfield{author}{\bibinfo{person}{Deqiang Li} {and} \bibinfo{person}{Qianmu
  Li}.} \bibinfo{year}{2020}\natexlab{}.
\newblock \showarticletitle{Adversarial Deep Ensemble: Evasion Attacks and
  Defenses for Malware Detection}.
\newblock \bibinfo{journal}{\emph{{IEEE} Trans. Inf. Forensics Secur.}}
  (\bibinfo{year}{2020}).
\newblock


\bibitem[\protect\citeauthoryear{Li, Cheng, Wu, Yuan, Gao, Yuan, and Luo}{Li
  et~al\mbox{.}}{2023a}]%
        {DBLP:journals/corr/abs-2303-08509}
\bibfield{author}{\bibinfo{person}{Heng Li}, \bibinfo{person}{Zhang Cheng},
  \bibinfo{person}{Bang Wu}, \bibinfo{person}{Liheng Yuan},
  \bibinfo{person}{Cuiying Gao}, \bibinfo{person}{Wei Yuan}, {and}
  \bibinfo{person}{Xiapu Luo}.} \bibinfo{year}{2023}\natexlab{a}.
\newblock \showarticletitle{Black-box Adversarial Example Attack towards {FCG}
  Based Android Malware Detection under Incomplete Feature Information}.
\newblock \bibinfo{journal}{\emph{CoRR}} (\bibinfo{year}{2023}).
\newblock


\bibitem[\protect\citeauthoryear{Li, Shan, Wenger, Zhang, Zheng, and Zhao}{Li
  et~al\mbox{.}}{2022b}]%
        {DBLP:conf/uss/LiSWZ0Z22}
\bibfield{author}{\bibinfo{person}{Huiying Li}, \bibinfo{person}{Shawn Shan},
  \bibinfo{person}{Emily Wenger}, \bibinfo{person}{Jiayun Zhang},
  \bibinfo{person}{Haitao Zheng}, {and} \bibinfo{person}{Ben~Y. Zhao}.}
  \bibinfo{year}{2022}\natexlab{b}.
\newblock \showarticletitle{Blacklight: Scalable Defense for Neural Networks
  against Query-Based Black-Box Attacks}. In \bibinfo{booktitle}{\emph{{USENIX}
  Security Symposium}}.
\newblock


\bibitem[\protect\citeauthoryear{Li, Zhou, Yuan, Luo, Gao, and Chen}{Li
  et~al\mbox{.}}{2021}]%
        {DBLP:conf/www/LiZYLGC21}
\bibfield{author}{\bibinfo{person}{Heng Li}, \bibinfo{person}{ShiYao Zhou},
  \bibinfo{person}{Wei Yuan}, \bibinfo{person}{Xiapu Luo},
  \bibinfo{person}{Cuiying Gao}, {and} \bibinfo{person}{Shuiyan Chen}.}
  \bibinfo{year}{2021}\natexlab{}.
\newblock \showarticletitle{Robust Android Malware Detection against
  Adversarial Example Attacks}. In \bibinfo{booktitle}{\emph{{WWW}}}.
\newblock


\bibitem[\protect\citeauthoryear{Li, Ji, Du, Li, and Wang}{Li
  et~al\mbox{.}}{2019}]%
        {DBLP:conf/ndss/LiJDLW19}
\bibfield{author}{\bibinfo{person}{Jinfeng Li}, \bibinfo{person}{Shouling Ji},
  \bibinfo{person}{Tianyu Du}, \bibinfo{person}{Bo Li}, {and}
  \bibinfo{person}{Ting Wang}.} \bibinfo{year}{2019}\natexlab{}.
\newblock \showarticletitle{TextBugger: Generating Adversarial Text Against
  Real-world Applications}. In \bibinfo{booktitle}{\emph{{NDSS}}}.
\newblock


\bibitem[\protect\citeauthoryear{Li, Xie, and Li}{Li et~al\mbox{.}}{2023b}]%
        {li2023sok}
\bibfield{author}{\bibinfo{person}{Linyi Li}, \bibinfo{person}{Tao Xie}, {and}
  \bibinfo{person}{Bo Li}.} \bibinfo{year}{2023}\natexlab{b}.
\newblock \showarticletitle{SoK: Certified Robustness for Deep Neural
  Networks}. In \bibinfo{booktitle}{\emph{{IEEE} S\&P}}.
\newblock


\bibitem[\protect\citeauthoryear{Liu, Tantithamthavorn, Li, and Liu}{Liu
  et~al\mbox{.}}{2023}]%
        {DBLP:journals/csur/LiuTLL23}
\bibfield{author}{\bibinfo{person}{Yue Liu}, \bibinfo{person}{Chakkrit
  Tantithamthavorn}, \bibinfo{person}{Li Li}, {and} \bibinfo{person}{Yepang
  Liu}.} \bibinfo{year}{2023}\natexlab{}.
\newblock \showarticletitle{Deep Learning for Android Malware Defenses: {A}
  Systematic Literature Review}.
\newblock \bibinfo{journal}{\emph{{ACM} Comput. Surv.}} (\bibinfo{year}{2023}).
\newblock


\bibitem[\protect\citeauthoryear{Mao, Fu, Wang, Ji, Zhang, Liu, Zhou, Liu,
  Beyah, and Wang}{Mao et~al\mbox{.}}{2022}]%
        {DBLP:conf/sp/MaoFWJ0LZLB022}
\bibfield{author}{\bibinfo{person}{Yuhao Mao}, \bibinfo{person}{Chong Fu},
  \bibinfo{person}{Saizhuo Wang}, \bibinfo{person}{Shouling Ji},
  \bibinfo{person}{Xuhong Zhang}, \bibinfo{person}{Zhenguang Liu},
  \bibinfo{person}{Jun Zhou}, \bibinfo{person}{Alex~X. Liu},
  \bibinfo{person}{Raheem Beyah}, {and} \bibinfo{person}{Ting Wang}.}
  \bibinfo{year}{2022}\natexlab{}.
\newblock \showarticletitle{Transfer Attacks Revisited: {A} Large-Scale
  Empirical Study in Real Computer Vision Settings}. In
  \bibinfo{booktitle}{\emph{{IEEE} S\&P}}.
\newblock


\bibitem[\protect\citeauthoryear{Mariconti, Onwuzurike, Andriotis, Cristofaro,
  Ross, and Stringhini}{Mariconti et~al\mbox{.}}{2017}]%
        {DBLP:conf/ndss/MaricontiOACRS17}
\bibfield{author}{\bibinfo{person}{Enrico Mariconti}, \bibinfo{person}{Lucky
  Onwuzurike}, \bibinfo{person}{Panagiotis Andriotis},
  \bibinfo{person}{Emiliano~De Cristofaro}, \bibinfo{person}{Gordon~J. Ross},
  {and} \bibinfo{person}{Gianluca Stringhini}.}
  \bibinfo{year}{2017}\natexlab{}.
\newblock \showarticletitle{MaMaDroid: Detecting Android Malware by Building
  Markov Chains of Behavioral Models}. In \bibinfo{booktitle}{\emph{{NDSS}}}.
\newblock


\bibitem[\protect\citeauthoryear{Ming, Xu, Wang, and Wu}{Ming
  et~al\mbox{.}}{2015}]%
        {DBLP:conf/ccs/MingXWW15}
\bibfield{author}{\bibinfo{person}{Jiang Ming}, \bibinfo{person}{Dongpeng Xu},
  \bibinfo{person}{Li Wang}, {and} \bibinfo{person}{Dinghao Wu}.}
  \bibinfo{year}{2015}\natexlab{}.
\newblock \showarticletitle{{LOOP:} Logic-Oriented Opaque Predicate Detection
  in Obfuscated Binary Code}. In \bibinfo{booktitle}{\emph{ACM {CCS}}}.
\newblock


\bibitem[\protect\citeauthoryear{Moser, Kruegel, and Kirda}{Moser
  et~al\mbox{.}}{2007}]%
        {DBLP:conf/acsac/MoserKK07}
\bibfield{author}{\bibinfo{person}{Andreas Moser}, \bibinfo{person}{Christopher
  Kruegel}, {and} \bibinfo{person}{Engin Kirda}.}
  \bibinfo{year}{2007}\natexlab{}.
\newblock \showarticletitle{Limits of Static Analysis for Malware Detection}.
  In \bibinfo{booktitle}{\emph{{ACSAC}}}.
\newblock


\bibitem[\protect\citeauthoryear{Murphy, Notkin, Griswold, and Lan}{Murphy
  et~al\mbox{.}}{1998}]%
        {DBLP:journals/tosem/MurphyNGL98}
\bibfield{author}{\bibinfo{person}{Gail~C. Murphy}, \bibinfo{person}{David
  Notkin}, \bibinfo{person}{William~G. Griswold}, {and}
  \bibinfo{person}{Erica~S.{-}C. Lan}.} \bibinfo{year}{1998}\natexlab{}.
\newblock \showarticletitle{An Empirical Study of Static Call Graph
  Extractors}.
\newblock \bibinfo{journal}{\emph{{ACM} Trans. Softw. Eng. Methodol.}}
  (\bibinfo{year}{1998}).
\newblock


\bibitem[\protect\citeauthoryear{Myles, Feudale, Liu, Woody, and Brown}{Myles
  et~al\mbox{.}}{2004}]%
        {myles2004introduction}
\bibfield{author}{\bibinfo{person}{Anthony~J Myles}, \bibinfo{person}{Robert~N
  Feudale}, \bibinfo{person}{Yang Liu}, \bibinfo{person}{Nathaniel~A Woody},
  {and} \bibinfo{person}{Steven~D Brown}.} \bibinfo{year}{2004}\natexlab{}.
\newblock \showarticletitle{An introduction to decision tree modeling}.
\newblock \bibinfo{journal}{\emph{Journal of Chemometrics: A Journal of the
  Chemometrics Society}} (\bibinfo{year}{2004}).
\newblock


\bibitem[\protect\citeauthoryear{Papernot, McDaniel, Goodfellow, Jha, Celik,
  and Swami}{Papernot et~al\mbox{.}}{2017}]%
        {DBLP:conf/ccs/PapernotMGJCS17}
\bibfield{author}{\bibinfo{person}{Nicolas Papernot},
  \bibinfo{person}{Patrick~D. McDaniel}, \bibinfo{person}{Ian~J. Goodfellow},
  \bibinfo{person}{Somesh Jha}, \bibinfo{person}{Z.~Berkay Celik}, {and}
  \bibinfo{person}{Ananthram Swami}.} \bibinfo{year}{2017}\natexlab{}.
\newblock \showarticletitle{Practical Black-Box Attacks against Machine
  Learning}. In \bibinfo{booktitle}{\emph{ACM AsiaCCS}}.
\newblock


\bibitem[\protect\citeauthoryear{Paszke, Gross, Massa, Lerer, Bradbury, Chanan,
  Killeen, Lin, Gimelshein, Antiga, Desmaison, K{\"{o}}pf, Yang, DeVito,
  Raison, Tejani, Chilamkurthy, Steiner, Fang, Bai, and Chintala}{Paszke
  et~al\mbox{.}}{2019}]%
        {DBLP:conf/nips/PaszkeGMLBCKLGA19}
\bibfield{author}{\bibinfo{person}{Adam Paszke}, \bibinfo{person}{Sam Gross},
  \bibinfo{person}{Francisco Massa}, \bibinfo{person}{Adam Lerer},
  \bibinfo{person}{James Bradbury}, \bibinfo{person}{Gregory Chanan},
  \bibinfo{person}{Trevor Killeen}, \bibinfo{person}{Zeming Lin},
  \bibinfo{person}{Natalia Gimelshein}, \bibinfo{person}{Luca Antiga},
  \bibinfo{person}{Alban Desmaison}, \bibinfo{person}{Andreas K{\"{o}}pf},
  \bibinfo{person}{Edward~Z. Yang}, \bibinfo{person}{Zachary DeVito},
  \bibinfo{person}{Martin Raison}, \bibinfo{person}{Alykhan Tejani},
  \bibinfo{person}{Sasank Chilamkurthy}, \bibinfo{person}{Benoit Steiner},
  \bibinfo{person}{Lu Fang}, \bibinfo{person}{Junjie Bai}, {and}
  \bibinfo{person}{Soumith Chintala}.} \bibinfo{year}{2019}\natexlab{}.
\newblock \showarticletitle{PyTorch: An Imperative Style, High-Performance Deep
  Learning Library}. In \bibinfo{booktitle}{\emph{NeurIPS}}.
\newblock


\bibitem[\protect\citeauthoryear{Pendlebury, Pierazzi, Jordaney, Kinder, and
  Cavallaro}{Pendlebury et~al\mbox{.}}{2019}]%
        {DBLP:conf/uss/PendleburyPJKC19}
\bibfield{author}{\bibinfo{person}{Feargus Pendlebury}, \bibinfo{person}{Fabio
  Pierazzi}, \bibinfo{person}{Roberto Jordaney}, \bibinfo{person}{Johannes
  Kinder}, {and} \bibinfo{person}{Lorenzo Cavallaro}.}
  \bibinfo{year}{2019}\natexlab{}.
\newblock \showarticletitle{{TESSERACT:} Eliminating Experimental Bias in
  Malware Classification across Space and Time}. In
  \bibinfo{booktitle}{\emph{{USENIX} Security Symposium}}.
\newblock


\bibitem[\protect\citeauthoryear{Pierazzi, Pendlebury, Cortellazzi, and
  Cavallaro}{Pierazzi et~al\mbox{.}}{2020}]%
        {DBLP:conf/sp/PierazziPCC20}
\bibfield{author}{\bibinfo{person}{Fabio Pierazzi}, \bibinfo{person}{Feargus
  Pendlebury}, \bibinfo{person}{Jacopo Cortellazzi}, {and}
  \bibinfo{person}{Lorenzo Cavallaro}.} \bibinfo{year}{2020}\natexlab{}.
\newblock \showarticletitle{Intriguing Properties of Adversarial {ML} Attacks
  in the Problem Space}. In \bibinfo{booktitle}{\emph{{IEEE} Symposium on
  S\&P}}.
\newblock


\bibitem[\protect\citeauthoryear{Preda, Madou, Bosschere, and Giacobazzi}{Preda
  et~al\mbox{.}}{2006}]%
        {DBLP:conf/amast/PredaMBG06}
\bibfield{author}{\bibinfo{person}{Mila~Dalla Preda}, \bibinfo{person}{Matias
  Madou}, \bibinfo{person}{Koen~De Bosschere}, {and} \bibinfo{person}{Roberto
  Giacobazzi}.} \bibinfo{year}{2006}\natexlab{}.
\newblock \showarticletitle{Opaque Predicates Detection by Abstract
  Interpretation}. In \bibinfo{booktitle}{\emph{AMAST}}.
\newblock


\bibitem[\protect\citeauthoryear{Quinlan}{Quinlan}{1996}]%
        {DBLP:journals/csur/Quinlan96}
\bibfield{author}{\bibinfo{person}{J.~Ross Quinlan}.}
  \bibinfo{year}{1996}\natexlab{}.
\newblock \showarticletitle{Learning Decision Tree Classifiers}.
\newblock \bibinfo{journal}{\emph{{ACM} Comput. Surv.}} (\bibinfo{year}{1996}).
\newblock


\bibitem[\protect\citeauthoryear{Salis, Sotiropoulos, Louridas, Spinellis, and
  Mitropoulos}{Salis et~al\mbox{.}}{2021}]%
        {DBLP:conf/icse/SalisSLSM21}
\bibfield{author}{\bibinfo{person}{Vitalis Salis}, \bibinfo{person}{Thodoris
  Sotiropoulos}, \bibinfo{person}{Panos Louridas}, \bibinfo{person}{Diomidis
  Spinellis}, {and} \bibinfo{person}{Dimitris Mitropoulos}.}
  \bibinfo{year}{2021}\natexlab{}.
\newblock \showarticletitle{PyCG: Practical Call Graph Generation in Python}.
  In \bibinfo{booktitle}{\emph{{ICSE}}}.
\newblock


\bibitem[\protect\citeauthoryear{Shen, Vervier, and Stringhini}{Shen
  et~al\mbox{.}}{2021}]%
        {DBLP:conf/ndss/ShenVS21}
\bibfield{author}{\bibinfo{person}{Yun Shen}, \bibinfo{person}{Pierre{-}Antoine
  Vervier}, {and} \bibinfo{person}{Gianluca Stringhini}.}
  \bibinfo{year}{2021}\natexlab{}.
\newblock \showarticletitle{Understanding Worldwide Private Information
  Collection on Android}. In \bibinfo{booktitle}{\emph{{NDSS}}}.
\newblock


\bibitem[\protect\citeauthoryear{Shen, Vervier, and Stringhini}{Shen
  et~al\mbox{.}}{2022}]%
        {DBLP:conf/uss/ShenVS22}
\bibfield{author}{\bibinfo{person}{Yun Shen}, \bibinfo{person}{Pierre~Antoine
  Vervier}, {and} \bibinfo{person}{Gianluca Stringhini}.}
  \bibinfo{year}{2022}\natexlab{}.
\newblock \showarticletitle{A Large-scale Temporal Measurement of Android
  Malicious Apps: Persistence, Migration, and Lessons Learned}. In
  \bibinfo{booktitle}{\emph{{USENIX} Security Symposium}}.
\newblock


\bibitem[\protect\citeauthoryear{Song, Li, Afroz, Garg, Kuznetsov, and
  Yin}{Song et~al\mbox{.}}{2022}]%
        {DBLP:conf/asiaccs/SongLAGKY22}
\bibfield{author}{\bibinfo{person}{Wei Song}, \bibinfo{person}{Xuezixiang Li},
  \bibinfo{person}{Sadia Afroz}, \bibinfo{person}{Deepali Garg},
  \bibinfo{person}{Dmitry Kuznetsov}, {and} \bibinfo{person}{Heng Yin}.}
  \bibinfo{year}{2022}\natexlab{}.
\newblock \showarticletitle{MAB-Malware: {A} Reinforcement Learning Framework
  for Blackbox Generation of Adversarial Malware}. In
  \bibinfo{booktitle}{\emph{ACM AsiaCCS}}.
\newblock


\bibitem[\protect\citeauthoryear{Statista}{Statista}{2023}]%
        {MobileMarket}
\bibfield{author}{\bibinfo{person}{Statista}.} \bibinfo{year}{2023}\natexlab{}.
\newblock \bibinfo{title}{{Mobile Operating Systems' Market}}.
\newblock
  \bibinfo{howpublished}{\url{https://www.statista.com/statistics/272698/global-market-share-held-by-mobile-operating-systems-since-2009/}}.
\newblock
\newblock
\shownote{[Accessed on Apr. 13, 2023].}


\bibitem[\protect\citeauthoryear{Suarez{-}Tangil and
  Stringhini}{Suarez{-}Tangil and Stringhini}{2022}]%
        {DBLP:journals/tdsc/Suarez-TangilS22}
\bibfield{author}{\bibinfo{person}{Guillermo Suarez{-}Tangil} {and}
  \bibinfo{person}{Gianluca Stringhini}.} \bibinfo{year}{2022}\natexlab{}.
\newblock \showarticletitle{Eight Years of Rider Measurement in the Android
  Malware Ecosystem}.
\newblock \bibinfo{journal}{\emph{{IEEE} Trans. Dependable Secur. Comput.}}
  (\bibinfo{year}{2022}).
\newblock


\bibitem[\protect\citeauthoryear{Suciu, Marginean, Kaya, III, and
  Dumitras}{Suciu et~al\mbox{.}}{2018}]%
        {DBLP:conf/uss/SuciuMKDD18}
\bibfield{author}{\bibinfo{person}{Octavian Suciu}, \bibinfo{person}{Radu
  Marginean}, \bibinfo{person}{Yigitcan Kaya}, \bibinfo{person}{Hal~Daum{\'{e}}
  III}, {and} \bibinfo{person}{Tudor Dumitras}.}
  \bibinfo{year}{2018}\natexlab{}.
\newblock \showarticletitle{When Does Machine Learning FAIL? Generalized
  Transferability for Evasion and Poisoning Attacks}. In
  \bibinfo{booktitle}{\emph{{USENIX} Security Symposium}}.
\newblock


\bibitem[\protect\citeauthoryear{Sun, Sun, Lu, and Mislove}{Sun
  et~al\mbox{.}}{2021}]%
        {DBLP:conf/uss/SunSLM21}
\bibfield{author}{\bibinfo{person}{Zhichuang Sun}, \bibinfo{person}{Ruimin
  Sun}, \bibinfo{person}{Long Lu}, {and} \bibinfo{person}{Alan Mislove}.}
  \bibinfo{year}{2021}\natexlab{}.
\newblock \showarticletitle{Mind Your Weight(s): {A} Large-scale Study on
  Insufficient Machine Learning Model Protection in Mobile Apps}. In
  \bibinfo{booktitle}{\emph{{USENIX} Security Symposium}}.
\newblock


\bibitem[\protect\citeauthoryear{Suya, Chi, Evans, and Tian}{Suya
  et~al\mbox{.}}{2020}]%
        {DBLP:conf/uss/Suya20}
\bibfield{author}{\bibinfo{person}{Fnu Suya}, \bibinfo{person}{Jianfeng Chi},
  \bibinfo{person}{David Evans}, {and} \bibinfo{person}{Yuan Tian}.}
  \bibinfo{year}{2020}\natexlab{}.
\newblock \showarticletitle{Hybrid Batch Attacks: Finding Black-box Adversarial
  Examples with Limited Queries}. In \bibinfo{booktitle}{\emph{USENIX Security
  Symposium}}.
\newblock


\bibitem[\protect\citeauthoryear{Szegedy, Zaremba, Sutskever, Bruna, Erhan,
  Goodfellow, and Fergus}{Szegedy et~al\mbox{.}}{2014}]%
        {DBLP:journals/corr/SzegedyZSBEGF13}
\bibfield{author}{\bibinfo{person}{Christian Szegedy},
  \bibinfo{person}{Wojciech Zaremba}, \bibinfo{person}{Ilya Sutskever},
  \bibinfo{person}{Joan Bruna}, \bibinfo{person}{Dumitru Erhan},
  \bibinfo{person}{Ian~J. Goodfellow}, {and} \bibinfo{person}{Rob Fergus}.}
  \bibinfo{year}{2014}\natexlab{}.
\newblock \showarticletitle{Intriguing properties of neural networks}. In
  \bibinfo{booktitle}{\emph{ICLR}}.
\newblock


\bibitem[\protect\citeauthoryear{Tanay and Griffin}{Tanay and Griffin}{2016}]%
        {DBLP:journals/corr/TanayG16}
\bibfield{author}{\bibinfo{person}{Thomas Tanay} {and}
  \bibinfo{person}{Lewis~D. Griffin}.} \bibinfo{year}{2016}\natexlab{}.
\newblock \showarticletitle{A Boundary Tilting Persepective on the Phenomenon
  of Adversarial Examples}.
\newblock \bibinfo{journal}{\emph{CoRR}} (\bibinfo{year}{2016}).
\newblock


\bibitem[\protect\citeauthoryear{Vall{\'{e}}e{-}Rai, Co, Gagnon, Hendren, Lam,
  and Sundaresan}{Vall{\'{e}}e{-}Rai et~al\mbox{.}}{1999}]%
        {DBLP:conf/cascon/Vallee-RaiCGHLS99}
\bibfield{author}{\bibinfo{person}{Raja Vall{\'{e}}e{-}Rai},
  \bibinfo{person}{Phong Co}, \bibinfo{person}{Etienne Gagnon},
  \bibinfo{person}{Laurie~J. Hendren}, \bibinfo{person}{Patrick Lam}, {and}
  \bibinfo{person}{Vijay Sundaresan}.} \bibinfo{year}{1999}\natexlab{}.
\newblock \showarticletitle{Soot - a Java bytecode optimization framework}. In
  \bibinfo{booktitle}{\emph{{CASCON}}}.
\newblock


\bibitem[\protect\citeauthoryear{VirusShare}{VirusShare}{2023}]%
        {virusshare}
\bibfield{author}{\bibinfo{person}{VirusShare}.}
  \bibinfo{year}{2023}\natexlab{}.
\newblock \bibinfo{title}{{VirusShare Dataset}}.
\newblock \bibinfo{howpublished}{\url{https://virusshare.com/}}.
\newblock
\newblock
\shownote{[Accessed on July 17, 2023].}


\bibitem[\protect\citeauthoryear{VirusTotal}{VirusTotal}{2023a}]%
        {VirusTotalAPI}
\bibfield{author}{\bibinfo{person}{VirusTotal}.}
  \bibinfo{year}{2023}\natexlab{a}.
\newblock \bibinfo{title}{{VirusTotal API Documentation}}.
\newblock
\newblock
\newblock
\shownote{[Accessed on Apr. 13, 2023].}


\bibitem[\protect\citeauthoryear{VirusTotal}{VirusTotal}{2023b}]%
        {VirusTotalSandbox}
\bibfield{author}{\bibinfo{person}{VirusTotal}.}
  \bibinfo{year}{2023}\natexlab{b}.
\newblock \bibinfo{title}{VirusTotal Sandboxes}.
\newblock
  \bibinfo{howpublished}{\url{https://support.virustotal.com/hc/en-us/articles/6253253596957}}.
\newblock
\newblock
\shownote{[Accessed on July 17, 2023].}


\bibitem[\protect\citeauthoryear{Vo, Abbasnejad, and Ranasinghe}{Vo
  et~al\mbox{.}}{2022}]%
        {DBLP:conf/ndss/VoAR22}
\bibfield{author}{\bibinfo{person}{Viet~Quoc Vo}, \bibinfo{person}{Ehsan
  Abbasnejad}, {and} \bibinfo{person}{Damith~C. Ranasinghe}.}
  \bibinfo{year}{2022}\natexlab{}.
\newblock \showarticletitle{RamBoAttack: {A} Robust and Query Efficient Deep
  Neural Network Decision Exploit}. In \bibinfo{booktitle}{\emph{{NDSS}}}.
\newblock


\bibitem[\protect\citeauthoryear{Wang, Sun, Dong, Li, Xue, Tyson, and Zhu}{Wang
  et~al\mbox{.}}{2021}]%
        {DBLP:journals/corr/abs-2111-10085}
\bibfield{author}{\bibinfo{person}{Wei Wang}, \bibinfo{person}{Ruoxi Sun},
  \bibinfo{person}{Tian Dong}, \bibinfo{person}{Shaofeng Li},
  \bibinfo{person}{Minhui Xue}, \bibinfo{person}{Gareth Tyson}, {and}
  \bibinfo{person}{Haojin Zhu}.} \bibinfo{year}{2021}\natexlab{}.
\newblock \showarticletitle{Exposing Weaknesses of Malware Detectors with
  Explainability-Guided Evasion Attacks}.
\newblock \bibinfo{journal}{\emph{CoRR}} (\bibinfo{year}{2021}).
\newblock


\bibitem[\protect\citeauthoryear{Wu, Chen, Gao, Fan, Liu, Wen, and Lyu}{Wu
  et~al\mbox{.}}{2021a}]%
        {DBLP:journals/tosem/WuCGFLWL21}
\bibfield{author}{\bibinfo{person}{Bozhi Wu}, \bibinfo{person}{Sen Chen},
  \bibinfo{person}{Cuiyun Gao}, \bibinfo{person}{Lingling Fan},
  \bibinfo{person}{Yang Liu}, \bibinfo{person}{Weiping Wen}, {and}
  \bibinfo{person}{Michael~R. Lyu}.} \bibinfo{year}{2021}\natexlab{a}.
\newblock \showarticletitle{Why an Android App Is Classified as Malware: Toward
  Malware Classification Interpretation}.
\newblock \bibinfo{journal}{\emph{{ACM} Trans. Softw. Eng. Methodol.}}
  (\bibinfo{year}{2021}).
\newblock


\bibitem[\protect\citeauthoryear{Wu, Guo, Wei, and Xing}{Wu
  et~al\mbox{.}}{2021b}]%
        {DBLP:conf/uss/Wu0WX21}
\bibfield{author}{\bibinfo{person}{Xian Wu}, \bibinfo{person}{Wenbo Guo},
  \bibinfo{person}{Hua Wei}, {and} \bibinfo{person}{Xinyu Xing}.}
  \bibinfo{year}{2021}\natexlab{b}.
\newblock \showarticletitle{Adversarial Policy Training against Deep
  Reinforcement Learning}. In \bibinfo{booktitle}{\emph{{USENIX} Security
  Symposium}}.
\newblock


\bibitem[\protect\citeauthoryear{Yang, Guo, Hao, Ciptadi, Ahmadzadeh, Xing, and
  Wang}{Yang et~al\mbox{.}}{2021}]%
        {DBLP:conf/uss/Yang0HCAX021}
\bibfield{author}{\bibinfo{person}{Limin Yang}, \bibinfo{person}{Wenbo Guo},
  \bibinfo{person}{Qingying Hao}, \bibinfo{person}{Arridhana Ciptadi},
  \bibinfo{person}{Ali Ahmadzadeh}, \bibinfo{person}{Xinyu Xing}, {and}
  \bibinfo{person}{Gang Wang}.} \bibinfo{year}{2021}\natexlab{}.
\newblock \showarticletitle{{CADE:} Detecting and Explaining Concept Drift
  Samples for Security Applications}. In \bibinfo{booktitle}{\emph{{USENIX}
  Security Symposium}}.
\newblock


\bibitem[\protect\citeauthoryear{Yang, Kong, Xie, and Gunter}{Yang
  et~al\mbox{.}}{2017}]%
        {DBLP:conf/acsac/YangK0G17}
\bibfield{author}{\bibinfo{person}{Wei Yang}, \bibinfo{person}{Deguang Kong},
  \bibinfo{person}{Tao Xie}, {and} \bibinfo{person}{Carl~A. Gunter}.}
  \bibinfo{year}{2017}\natexlab{}.
\newblock \showarticletitle{Malware Detection in Adversarial Settings:
  Exploiting Feature Evolutions and Confusions in Android Apps}. In
  \bibinfo{booktitle}{\emph{{ACSAC}}}.
\newblock


\bibitem[\protect\citeauthoryear{Yang, Zhao, Wang, Zhang, Li, Pei, Karlaš,
  Liu, Guo, Zhang, and Li}{Yang et~al\mbox{.}}{2022}]%
        {YangImp2022}
\bibfield{author}{\bibinfo{person}{Zhuolin Yang}, \bibinfo{person}{Zhikuan
  Zhao}, \bibinfo{person}{Boxin Wang}, \bibinfo{person}{Jiawei Zhang},
  \bibinfo{person}{Linyi Li}, \bibinfo{person}{Hengzhi Pei},
  \bibinfo{person}{Bojan Karlaš}, \bibinfo{person}{Ji Liu},
  \bibinfo{person}{Heng Guo}, \bibinfo{person}{Ce Zhang}, {and}
  \bibinfo{person}{Bo Li}.} \bibinfo{year}{2022}\natexlab{}.
\newblock \showarticletitle{Improving Certified Robustness via Statistical
  Learning with Logical Reasoning}. In \bibinfo{booktitle}{\emph{NeurIPS}}.
\newblock


\bibitem[\protect\citeauthoryear{Zhang, Zhang, Liu, Wang, Diao, and Guo}{Zhang
  et~al\mbox{.}}{2021}]%
        {DBLP:conf/icpads/ZhangZLWDG21}
\bibfield{author}{\bibinfo{person}{Jin Zhang}, \bibinfo{person}{Chennan Zhang},
  \bibinfo{person}{Xiangyu Liu}, \bibinfo{person}{Yuncheng Wang},
  \bibinfo{person}{Wenrui Diao}, {and} \bibinfo{person}{Shanqing Guo}.}
  \bibinfo{year}{2021}\natexlab{}.
\newblock \showarticletitle{ShadowDroid: Practical Black-box Attack against
  ML-based Android Malware Detection}. In \bibinfo{booktitle}{\emph{{ICPADS}}}.
\newblock


\bibitem[\protect\citeauthoryear{Zhang, Wang, Shen, Ji, Luo, and Wang}{Zhang
  et~al\mbox{.}}{2020a}]%
        {DBLP:conf/uss/ZhangWSJL020}
\bibfield{author}{\bibinfo{person}{Xinyang Zhang}, \bibinfo{person}{Ningfei
  Wang}, \bibinfo{person}{Hua Shen}, \bibinfo{person}{Shouling Ji},
  \bibinfo{person}{Xiapu Luo}, {and} \bibinfo{person}{Ting Wang}.}
  \bibinfo{year}{2020}\natexlab{a}.
\newblock \showarticletitle{Interpretable Deep Learning under Fire}. In
  \bibinfo{booktitle}{\emph{{USENIX} Security Symposium}}.
\newblock


\bibitem[\protect\citeauthoryear{Zhang, Zhang, Zhong, Ding, Cao, Zhang, Zhang,
  and Yang}{Zhang et~al\mbox{.}}{2020b}]%
        {DBLP:conf/ccs/ZhangZZDCZZY20}
\bibfield{author}{\bibinfo{person}{Xiaohan Zhang}, \bibinfo{person}{Yuan
  Zhang}, \bibinfo{person}{Ming Zhong}, \bibinfo{person}{Daizong Ding},
  \bibinfo{person}{Yinzhi Cao}, \bibinfo{person}{Yukun Zhang},
  \bibinfo{person}{Mi Zhang}, {and} \bibinfo{person}{Min Yang}.}
  \bibinfo{year}{2020}\natexlab{b}.
\newblock \showarticletitle{Enhancing State-of-the-art Classifiers with {API}
  Semantics to Detect Evolved Android Malware}. In
  \bibinfo{booktitle}{\emph{{ACM CCS}}}.
\newblock


\bibitem[\protect\citeauthoryear{Zhao, Zhou, Zhu, Zhan, Zhou, Li, Yu, Yuan, and
  Luo}{Zhao et~al\mbox{.}}{2021}]%
        {DBLP:conf/ccs/ZhaoZZZZLYYL21}
\bibfield{author}{\bibinfo{person}{Kaifa Zhao}, \bibinfo{person}{Hao Zhou},
  \bibinfo{person}{Yulin Zhu}, \bibinfo{person}{Xian Zhan},
  \bibinfo{person}{Kai Zhou}, \bibinfo{person}{Jianfeng Li},
  \bibinfo{person}{Le Yu}, \bibinfo{person}{Wei Yuan}, {and}
  \bibinfo{person}{Xiapu Luo}.} \bibinfo{year}{2021}\natexlab{}.
\newblock \showarticletitle{Structural Attack against Graph Based Android
  Malware Detection}. In \bibinfo{booktitle}{\emph{ACM {CCS}}}.
\newblock


\end{thebibliography}

\appendix

\newpage

\section{APK File Structure}
\label{appendix:apk}

\begin{figure}[t]   
	\centering  
	\includegraphics[width=1\linewidth]{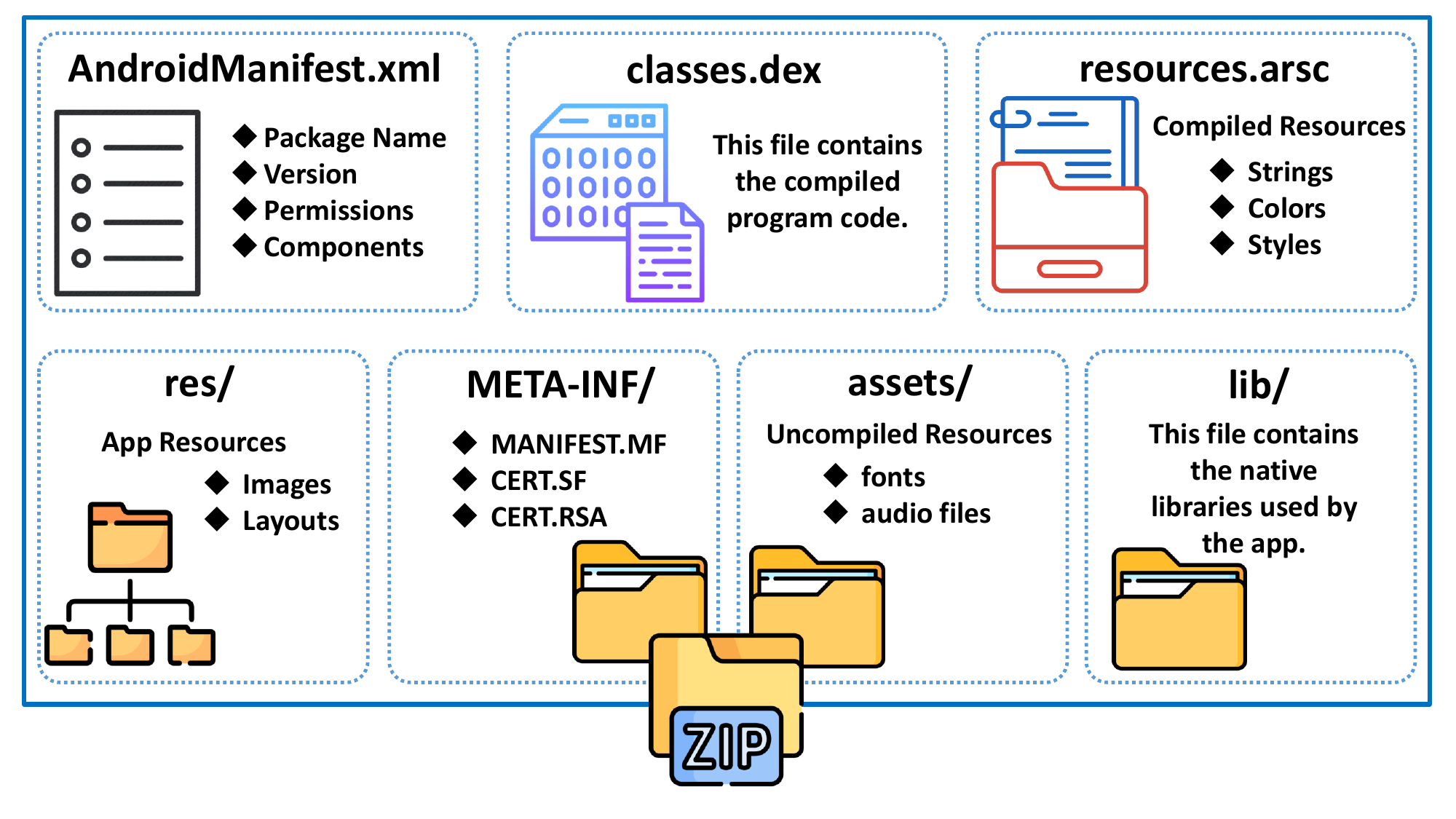}
	\caption{The APK file structure is a zip archive containing all the resources, assets, and compiled code needed for an application to run on an Android device.}  
	\label{fig:apk}
\end{figure}

As shown in Figure~\ref{fig:apk}, the APK file structure is designed to encapsulate all the necessary components of an application, including code, resources, and metadata.
It usually contains the following key components: \textit{AndroidManifest.xml} is a crucial XML file of an APK file.
as it contains essential information about the application, such as its package name, version, permissions, and list of components (e.g., activities, services, broadcast receivers, and content providers).
The Android system uses the file to manage and execute the application correctly.
\textit{classes.dex} is the DEX file containing the compiled bytecode of the application.
It is executed by the Android Runtime (ART) or the Dalvik Virtual Machine (DVM).
\textit{resources.arsc} contains compiled resources, such as strings, colors, and styles, used by the application.
\textit{res/} is the resources directory containing various files needed for the application's user interface, such as images, XML layouts, and string values.
\textit{assets/} contains any additional files the application needs, such as fonts, audio files, or other binary data
\textit{META-INF/} contains metadata about the APK file, such as digital signatures and certificate information.
\textit{lib/} contains the compiled libraries if an application uses native code written in C or C++.
The APK file is essentially a compressed ZIP archive, which bundles all the above components together.

\section{Four ML-based AMD Methods}
\label{appendix:mlamd}

\paragraphbe{Drebin}
Drebin~\cite{DBLP:conf/ndss/ArpSHGR14} takes into account a broad spectrum of features derived from both the manifest file and DEX codes.
From the manifest file, it extracts four sets of string features, including hardware components, requested permissions, app components, and filtered intents.
For the DEX codes, Drebin disassembles the \textit{classes.dex} file into smali files and extracts an additional four sets of string features: restricted API calls,  permissions, suspicious API calls, and network addresses.
Subsequently, it obtains these eight feature sets from the training dataset and merges them into a binary feature vector.
Drebin employs the Linear SVM classifier to discern patterns between benign applications and malware.
Moreover, Drebin offers explanations for its decisions by utilizing the weights and detection scores of the Linear SVM classifier.

\paragraphbe{Drebin-DL}
Drebin-DL~\cite{DBLP:conf/esorics/GrossePMBM17} employs the same feature set as Drebin, but utilizes MLP as the classifier.
Despite sharing the same feature space, Drebin-DL achieves superior Android malware detection performance compared to Drebin.\footnote{Though the primary focus of Drebin-DL~\cite{DBLP:conf/esorics/GrossePMBM17} is on model robustness, it also introduces an improved AMD method.}

\paragraphbe{MaMadroid}
MaMadroid~\cite{DBLP:conf/ndss/MaricontiOACRS17} focuses on the function call graph of Android applications to detect malware.
By extracting the FCG, It is able to analyze the behavioral patterns of applications, thereby providing valuable insights for malware detection.
To enhance robustness against API changes in the Android framework, it abstracts functions into different states based on package or family names, creating a higher level of abstraction less prone to minor changes or variations.
Subsequently, it constructs a Markov chain model that captures the transition probabilities between states, i.e., families or packages of target functions.
This model helps represent the dynamic behavior of applications, thus allowing the method to identify distinctive patterns exhibited by malicious applications.
Finally, it trains a machine learning classifier, such as RF, SVM, or kNN, to detect malware.

\paragraphbe{APIGraph}
Distinct from the aforementioned AMD methods, APIGraph~\cite{DBLP:conf/ccs/ZhangZZDCZZY20} is a general framework designed to enhance AMD methods that use the function call-related features.
Initially, APIGraph constructs a comprehensive knowledge graph about APIs by extracting API relationships from the Android API documentation.
This knowledge graph provides an organized and structured representation of the API ecosystem, enabling a deeper understanding of the connections between different APIs.
To enhance the representation ability of function call-related features and better capture the underlying semantics, APIGraph employs clustering algorithms to aggregate functions based on their relationships within the knowledge graph.
When applied to specific AMD methods, APIGraph replaces individual functions with the corresponding cluster to which the function belongs.
This replacement provides a higher level of abstraction, reducing the impact of minor variations and potential obfuscation techniques in the analyzed applications.
For example, to augment Drebin, APIGraph substitutes the binary vector of API occurrence with the one representing API cluster occurrence.

\section{Framework Algorithm}
\label{appendix:algorithm}

The primary procedure of \system is shown in Algorithm~\ref{alg:overview}.
Within this algorithm, $g$ represents the target classifier, $N$ stands for the query budget, $\mathcal{D}$ symbolizes the Android documentation, $\mathbb{B}$ signifies the set of randomly sampled benign applications, and $\mathbb{P}$ denotes the malware perturbation set.

\system first builds the malware perturbation set from the Android documentation and the randomly sampled benign applications (lines 1).
Then \system organizes the perturbations within the perturbation set into the perturbation selection tree (line 2).
For a malicious application, \system initially obtains the initial malicious confidence (line 4).
Throughout each iteration, \system samples a perturbation from the perturbation selection tree (line 6) and implements the perturbation within the APK (line 7).
Subsequently, the perturbed malware is submitted to the target classifier (line 8).
In the event that the target classifier is caused to misclassify, the attack is deemed successful; otherwise, \system proceeds to adjust the perturbation selection tree (lines 9-13).
Perturbations that result in increasing malware confidence are not recorded by the framework (lines 14-17).

\begin{algorithm}[t]
    \footnotesize
    \caption{\system}
    \label{alg:overview}
    \begin{algorithmic}[1]
        
        \Require{
        Target classifier $g$;
        Target malicious application $z$;
        Query budget $N$;
        Android documentation $\mathcal{D}$;
        Benign applications $\mathbb{B}$.
        }
        
        \Ensure{
        Adversarial application $z'$.}
        
        \Statex
        \State $\mathbb{P} \leftarrow $ BuildPerturbation($\mathcal{D}$, $\mathbb{B}$) \Comment{Build the malware perturbation set.}
        \State $\mathbb{T} \leftarrow $ InitTree($\mathbb{P}$) \Comment{Develop the perturbation selection tree}
        \State $q \leftarrow 0$ \Comment{$q$ represents the query times.}
        \State $y \leftarrow g(z)$ \Comment{Obtain the initial malicious confidence.}
        \While {$q \leq N$}
            \State $P \leftarrow $ SelectPerturbation($\mathbb{T}$) \Comment{Sample the perturbation node $P$.}
            \State $z' \leftarrow$ Implement($P$, $z$) \Comment{Add the perturbation in problem space.}
            \State $y' \leftarrow g(z')$ \Comment{Obtain the malicious confidence.}
            \If {$y'$ is benign}
                \State \Return $z'$; \Comment{Attack successful.}
            \Else 
                \State $\mathbb{T} \leftarrow $ AdjustTree($\mathbb{T}$, $P$, $y$, $y'$) \Comment{Apply the adjustment policy.}
            \EndIf
            \If {$y' \leq y $} \Comment{Apply the evasive perturbation.}
                \State $z \leftarrow z'$
                \State $y \leftarrow y'$
            \EndIf
            \State $q \leftarrow q + 1$
        \EndWhile
        \State \Return Failure
        
    \end{algorithmic}
\end{algorithm}

\section{Intuition Validation}
\label{appendix:validation}

Our assumption that perturbations sharing semantic similarity are more likely to demonstrate equivalent evasion effectiveness is rooted in observations of benign Android applications.
Such applications often utilize semantically similar manifest elements to facilitate specific functionalities.
For example, a music-related application may necessitate audio capabilities for a device and declare both a \textit{uses-feature} element with \textit{android.hardware.audio.output} value and a \textit{uses-feature} element with \textit{android.hardware.audio.pro} value.
Consequently, models trained on such data will recognize these patterns suggesting that both values contribute to the classification of the application as benign.
Moreover, the only difference between \system and the baseline methods, i.e., MAB and RA, resides in the perturbation selection strategy, while the malware perturbation set remains the same across all methods.
Therefore, the enhanced performance of \system can be attributed to this assumption to some extent.

To provide empirical validation of our assumption, we evaluate the effectiveness of the semantically related perturbations in our implemented ML-based AMD methods.
To be specific, we randomly sample 100 malicious samples detected by the ML-based AMD methods in the test set in the primary dataset.
Then, we randomly sample 100 pairs of semantically related perturbations from the malware perturbation set.
For every pair of semantically related perturbation pair, we individually implement each perturbation in a pair into the malware sample and observe if they yield similar results, e.g., both of them decrease the confidence of the malware label.
Through this analysis, we find that 64.18\% of semantically related perturbation pairs against Drebin and 67.69\% against APIGraph yield similar rewards.
These results suggest that semantically related perturbations likely possess similar evasion effectiveness, thereby supporting our assumption.

\section{Malware Manipulation Requirements}
\label{appendix:manipulation}

In the context of malware, manipulations must adhere to the following requirements.

\paragraphbe{R1: All-features Influence}
Given that the target malware detectors are entirely unknown, manipulations should be capable of affecting different feature categories (e.g., manifest-related feature and code-related feature).

\paragraphbe{R2: Functional Consistency}
The functionality of the manipulated malware should maintain its malicious functional consistency before and after the manipulation.

\paragraphbe{R3: Robust to Pre-Processing}
Manipulations in the problem space should exhibit robustness against pre-processing techniques.
This means they should not be eliminated by static analysis inspection \cite{DBLP:conf/acsac/MoserKK07,DBLP:journals/tdsc/DemontisMBMARCG19}.

Existing manipulation methods are summarized below.

\paragraphbe{Inserting Dead Codes}
To preserve functional consistency, Chen \textit{et al}. \cite{DBLP:journals/tifs/0002LWW0N0020} insert dead codes (e.g., no-op calls) into smali files, while Bostani \textit{et al}. \cite{DBLP:journals/corr/abs-2110-03301} insert code gadgets into unreachable program positions (e.g., after return statements).
However, these codes could be easily detected and eliminated, violating \textbf{R3}.
For instance, Karer \textit{et al}. \cite{7475289} propose an SSA-based static analysis method for dead code elimination.

\paragraphbe{Opaque Predicates}
Pierazzi \textit{et al}. \cite{DBLP:conf/sp/PierazziPCC20} devise a method called opaque predicates to insert new API calls.
Specifically, opaque predicates are conditional statements that are practically impossible to be true but difficult to determine for static analysis.
However, it is straightforward to approximate the probability of the truth of these conditional statements.
By doing so, static analysis tools may remove the opaque predicates~\cite{DBLP:conf/ccs/MingXWW15,DBLP:conf/amast/PredaMBG06}, causing them to violate \textbf{R3}.

\paragraphbe{Try-Catch Wrappers}
Li \textit{et al}. \cite{DBLP:journals/corr/abs-2303-08509} employ try-catch blocks to wrap new function calls to evade malware detection.
To maintain \textbf{R2}, this method inserts try-catch blocks with statements that always trigger exceptions.
New function calls after the exception-triggering statements will never be executed due to the early exceptions.
Hence, it is possible for static analysis to detect such dead code by identifying the exceptions.
Moreover, this method may introduce extra features (e.g., the abuse of try-catch statements), which could be used to reveal malicious APKs.
Therefore, this method violates \textbf{R3}.

\section{Manipulation Examples}
\label{appendix:manifest}

\begin{figure}[t]
\label{list:sourceapk}
\begin{lstlisting}[language=XML, caption=Orginal AndroidManifest.xml of malicious APK]
<?xml version="1.0" encoding="utf-8" standalone="no"?><manifest xmlns:android="http://schemas.android.com/apk/res/android" package="com.lifeapps.modernwindowdesign" platformBuildVersionCode="22" platformBuildVersionName="5.1.1-1819727">
    <uses-permission android:name="android.permission.ACCESS_NETWORK_STATE"/>
    <uses-permission android:name="android.permission.ACCESS_FINE_LOCATION"/>
    <uses-permission android:name="android.permission.CAMERA"/>
    <uses-permission android:name="android.permission.READ_EXTERNAL_STORAGE"/>
    <uses-permission android:name="android.permission.WRITE_EXTERNAL_STORAGE"/>
    <uses-permission android:name="android.permission.INTERNET"/>
    <application android:allowBackup="true" android:icon="@drawable/ic_launcher" android:label="@string/app_name" android:theme="@style/AppTheme">
        <activity android:configChanges="keyboardHidden|orientation|screenSize" android:hardwareAccelerated="true" android:label="@string/app_name" android:name="com.lifeapps.modernwindowdesign.MainActivity" android:screenOrientation="user">
            <intent-filter>
                <action android:name="android.intent.action.MAIN"/>
                <category android:name="android.intent.category.LAUNCHER"/>
            </intent-filter>
        </activity>
        <activity android:configChanges="keyboard|keyboardHidden|orientation|screenLayout|screenSize|smallestScreenSize|uiMode" android:name="com.google.android.gms.ads.AdActivity"/>
        <meta-data android:name="com.google.android.gms.version" android:value="@integer/google_play_services_version"/>
    </application>
</manifest>
\end{lstlisting}
\end{figure}

\begin{figure}[t]
    \label{list:advapk}
    \begin{lstlisting}[language=XML, caption=Perturbed AndroidManifest.xml of malicious APK]
<?xml version="1.0" encoding="utf-8" standalone="no"?><manifest xmlns:android="http://schemas.android.com/apk/res/android" package="com.lifeapps.modernwindowdesign" platformBuildVersionCode="22" platformBuildVersionName="5.1.1-1819727">
    <uses-permission android:name="android.permission.ACCESS_NETWORK_STATE"/>
    <uses-permission android:name="android.permission.ACCESS_FINE_LOCATION"/>
    <uses-permission android:name="android.permission.CAMERA"/>
    <uses-permission android:name="android.permission.READ_EXTERNAL_STORAGE"/>
    <uses-permission android:name="android.permission.WRITE_EXTERNAL_STORAGE"/>
    <uses-permission android:name="android.permission.INTERNET"/>
    <application android:allowBackup="true" android:icon="@drawable/ic_launcher" android:label="@string/app_name" android:theme="@style/AppTheme">
        <activity android:configChanges="keyboardHidden|orientation|screenSize" android:hardwareAccelerated="true" android:label="@string/app_name" android:name="com.lifeapps.modernwindowdesign.MainActivity" android:screenOrientation="user">
            <intent-filter>
                <action android:name="android.intent.action.MAIN"/>
                <category android:name="android.intent.category.LAUNCHER"/>
            </intent-filter>
        </activity>
        <activity android:configChanges="keyboard|keyboardHidden|orientation|screenLayout|screenSize|smallestScreenSize|uiMode" android:name="com.google.android.gms.ads.AdActivity"/>
        <meta-data android:name="com.google.android.gms.version" android:value="@integer/google_play_services_version"/>
        <receiver android:exported="true" android:name="com.seattleclouds.scm.PushManagerReceiver" android:process=":sojVFdeGbnNtwndwghJH">
            <intent-filter>
                <action android:name="com.gnip.RRpEgRJXMCYKeff"/>
            </intent-filter>
        </receiver>
        <provider android:authorities="com.gnip.iGAMPzaW" android:enabled="true" android:exported="true" android:name="com.seattleclouds.util.InternalFileContentProvider" android:process=":LOPuTPUUlLRagJmqNRnj"/>
        <activity android:exported="true" android:name="coma.sbda.uygi.Vctq">
            <intent-filter>
                <action android:name="JcqPliYQ"/>
                <category android:name="android.intent.category.INFO"/>
            </intent-filter>
        </activity>
    </application>
    <uses-feature android:name="android.software.managed_users"/>
    <uses-permission android:name="android.permission.USE_ICC_AUTH_WITH_DEVICE_IDENTIFIER"/>
</manifest>
    \end{lstlisting}
\end{figure}

We employ adversarial Android malware generated by \system against the VirusTotal to illustrate the manipulations performed.\footnote{The VirusTotal analysis link of the original APK at \url{https://www.virustotal.com/gui/file/094282e6c9a78395787c50f0880759c44bb16a7394e852a24b7b4ad63d9b5fb7}.
The VirusTotal analysis link of the adversarial APK at \url{https://www.virustotal.com/gui/file/5558470e0bcfd3c137921aa49f3d5a92c7f513ef4aa418c96237544f90841b82}.}
As a result, there are 15 antivirus engines that detect the original APK, but only 7 antivirus engines can detect the adversarial APK.

We utilize Apktool to extract the \textit{AndroidManifest.xml} of the two APKs.
Listing 1 presents the \textit{AndroidManifest.xml} of the original APK, and Listing 2 depicts the adversarial APK.
Notably, the code after line 17 in Listing 2 has been added by \system.
The modifications introduced by \system include the injection of a \textit{uses-feature} element, a \textit{uses-permission} element, and a \textit{category} element.
Additionally, \system injects a broadcast receiver component and a content provider component.

\section{Perturbation Clustering}
\label{appendix:cluster}

As discussed in Section~\ref{sec:overview}, semantically related perturbations are more likely to yield similar rewards for an attack.
To capitalize on this characteristic, \system develops a customized clustering algorithm to group semantically related perturbations.
Based on the central insight that semantically related permissions or actions share similar text representations, the algorithm continuously merges the two most similar clusters according to their keyword similarity $K_{s}$ until no $K_{s}$ of existing cluster pairs surpasses the pre-defined threshold $T_{s}$.

As illustrated in Equation~\ref{equ:sim}, the keyword similarity of a cluster pair $c_{i}, c_{j}$ is calculated, where $\mathop{Overlapping}(c_{i}, c_{j})$ counts the number of identical keywords in the two clusters.
\begin{equation}
    \label{equ:sim}
    K_{s} = \frac{\mathop{Overlapping}(c_{i}, c_{j})}{\min(|c_{i}|, |c_{j}|)}.
\end{equation}

As depicted in Algorithm \ref{alg:clustering}, \system initializes every perturbation in the pool $P$ as a cluster.
Subsequently, \system computes the $K_{s}$ for every cluster pair (lines 4-5).
Then, the cluster pair with $K_{s}$ exceeding $T_{s}$ is merged (lines 6-9).
The process terminates when the perturbation groups no longer change (lines 11-13).

\begin{algorithm}[t]
    \footnotesize
    \caption{Perturbation Clustering}
    \label{alg:clustering}
    \begin{algorithmic}[1]
        
        \Require{
        Perturbation set $\mathbb{P}$;
        Keyword similarity threshold $T_s$
        }
        
        \Ensure{
        Perturbation groups $C'$.}
        
        \Statex
        \State $C \leftarrow $ Init($\mathbb{P}$) \Comment{Initialize the clustering group.}
        \State $C' \leftarrow C $
        \While {True}
            \For {$c_{i}, c_{j} \leftarrow $ Combinations($C$, 2)}
                \State $K_{s}$ = KeywordSim($c_{i}$, $c_{j}$)       \Comment{Compute the keyword-based similarity.}
                \If{$K_{s} > T_{s}$}
                    \State $C' \leftarrow $ MergeGroup($C$, $c_{i}$, $c_{j}$) \Comment{Merge the similar groups.}
                    \State \textbf{break}
                \EndIf
            \EndFor
            \If{$C = C'$}
                \State \textbf{break}
            \EndIf
        \EndWhile
        \State \Return $C'$
    \end{algorithmic}
\end{algorithm}

\begin{table}[t]
    \setlength{\abovecaptionskip}{0pt}
	\caption{Detection performance of the four ML-based AMD methods.}
 
	\begin{tabular}[centering,width=0.5*\linewidth]{@{}C{2cm}C{1cm}C{1cm}C{0.9cm}C{0.6cm}C{1.2cm}@{}}
		\toprule
            AMD Methods & Classifier & Precision & Recall & F1 & AUROC\\
		\midrule
        Drebin&SVM&0.73&0.88&0.80&0.96\\
        Drebin-DL&MLP&0.73&0.89&0.80&0.96\\
        APIGraph&SVM&0.73&0.88&0.80&0.96\\
        \multirow{2}{*}{MaMadroid}  &RF&0.76&0.80&0.78&0.94\\
                                    &3-NN&0.65&0.72&0.69&0.90\\
        \bottomrule
  	\end{tabular}
 
	\label{tab:AMDPerformance}
\end{table}

\section{Details of Target Model}
\label{appendix:targetmodel}

\paragraphbe{Implementation} For Drebin, we utilize the same implementation provided by Pierazzi \textit{et al}. \cite{DBLP:conf/sp/PierazziPCC20}.
For Drebin-DL, we employ the same feature extractor as Drebin and implement an MLP using PyTorch, following the descriptions in Grosse \textit{et al}. \cite{DBLP:conf/esorics/GrossePMBM17}.
As for APIGraph, we make use of its official code available at \url{https://github.com/seclab-fudan/APIGraph}.
Regarding MaMadroid, we re-implement it using Androguard \cite{Androguard}, adhering to the descriptions from the original paper and the configurations from the official code, which can be accessed at \url{https://bitbucket.org/gianluca_students/mamadroid_code/src/master/}.

\paragraphbe{Detection Performance} The TPR values for Drebin, Drebin-DL, APIGraph, MaMadroid with RF, and MaMadroid with 3-NN are 0.88, 0.89, 0.88, 0.80, and 0.72, respectively.
Supplementary measures such as precision, recall, f1 score, and AUROC for the four ML-based AMD techniques are detailed in Table~\ref{tab:AMDPerformance}.

\section{Baseline Methods}
\label{appendix:baseline}

\paragraphbe{MAB} In the original implementation, the MAB algorithm is designed specifically for Windows PE malware (source code available at: \url{https://github.com/weisong-ucr/MAB-malware}).
To adapt this baseline MAB algorithm for generating adversarial Android malware, we follow the descriptions and code configurations provided in the relevant literature.
Specifically, we treat the second layer in the perturbation selection tree as the arms of the multi-armed bandit machine.
We consider all leaf nodes that belong to the descendants of the corresponding arm as the contextual information for the bandit machine.
The detailed algorithm can be found in Algorithm~\ref{alg:MAB}.

\paragraphbe{RA} The random attack is to select perturbation from the perturbation space uniformly at random.
The detailed procedure is shown in Algorithm~\ref{alg:RA}.
RA uniformly samples the malware perturbation in the malware perturbation set iteratively, injects the selected perturbations, and queries the target model with the perturbed application until success is achieved or the query budget depletes.

To enable fair comparison, we utilize the same malware perturbation set between \system and the baseline methods.
The only difference between \system and the baseline methods, i.e., MAB and RA, lies in their approach to perturbation selection.
\system employs a perturbation selection tree, MAB utilizes the multi-armed bandit algorithm, and RA makes random, indiscriminate perturbation selections.
Consequently, the enhanced performance of \system can be ascribed to our strategic perturbation selection, while results of RA serve as a testament to our perturbation set.

\begin{algorithm}[t]
    \footnotesize
    \caption{MAB}
    \label{alg:MAB}
    \begin{algorithmic}[1]
        
        \Require{
        Target classifier $g$;
        Target malicious application $z$;
        Query budget $N$;
        Malware perturbation set $\mathbb{P}$.
        }
        
        \Ensure{
        Adversarial application $z'$.}
        
        \Statex
        \State $\mathbb{B} \leftarrow$ BuildBandit($\mathbb{P}$) \Comment{Build the multi-arm bandit machine $\mathbb{P}$.}
        \State $q \leftarrow 0$ \Comment{$q$ represents the query times.}
        \While {$q \leq N$}
            \State $P \leftarrow$ SelectPerturbation($\mathbb{B}$) \Comment{Obtain the perturbation $P$.}
            \State $z' \leftarrow$ Implement($P$, $z$) \Comment{Add the perturbation in problem space.}
            \State $y \leftarrow g(z')$ \Comment{Obtain the model feedback.}
            \If {$y$ is benign}
                \State \Return $z'$; \Comment{Attack successful.}
            \Else 
                \State $\mathbb{B} \leftarrow $ UpdateBandit($\mathbb{B}$, $P$, $y$) \Comment{Update the bandit machine.}
            \EndIf
            \State $q \leftarrow q + 1$
        \EndWhile
        \State \Return Failure
        
    \end{algorithmic}
\end{algorithm}

\begin{algorithm}[t]
    \footnotesize
    \caption{Random Attack}
    \label{alg:RA}
    \begin{algorithmic}[1]
        
        \Require{
        Target classifier $g$;
        Target malicious application $z$;
        Query budget $N$;
        Malware perturbation set $\mathbb{P}$.
        }
        
        \Ensure{
        Adversarial application $z'$.}
        
        \Statex
        \State $q \leftarrow 0$ \Comment{$q$ represents the query times.}
        \While {$q \leq N$}
            \State $P \leftarrow$ RandomSample($\mathbb{P}$) \Comment{Random sample uniformly the perturbation $P$.}
            \State $z' \leftarrow$ Implement($P$, $z$) \Comment{Add the perturbation in problem space.}
            \State $y \leftarrow g(z')$ \Comment{Obtain the model feedback.}
            \If {$y$ is benign}
                \State \Return $z'$; \Comment{Attack successful.}
            \EndIf
            \State $q \leftarrow q + 1$
        \EndWhile
        \State \Return Failure
        
    \end{algorithmic}
\end{algorithm}

\section{Actual Number of Successfully Generated Adversarial Android Malware}
\label{appendix:actualnumber}

Table~\ref{tab:attackeffectActualNumber} provides the actual number of adversarial applications successfully generated by \system that can evade the detection by the targeted ML-based AMD method.
Recognizing the bugs and corner cases present in the FlowDroid research prototype, as discussed by Pierazzi \textit{et al}.~\cite{DBLP:conf/sp/PierazziPCC20}, we encounter crashes during the modification of APKs with FlowDroid.
Following the previous work~\cite{DBLP:conf/sp/PierazziPCC20}, these crashes are not indicative of limitations in our method.
Hence such instances are excluded from our ASR computation.

\begin{table*}[t]\small\centering
	\setlength{\abovecaptionskip}{0pt}
	\caption{The number of adversarial Android malware successfully generated by \system, capable of evading the ML-based AMD method, selected from a random sample of 100 malware.}
 
	\begin{tabular}{@{}C{2.5cm}C{1.8cm}C{1.7cm}C{1.7cm}C{1.7cm}C{1.7cm}C{1.7cm}@{}}
		\toprule
		\multirow{5}{*}{\begin{tabular}[c]{@{}c@{}}Attack Methods\end{tabular}} &
		\multirow{5}{*}{Query Budgets}                                                           &
		\multicolumn{5}{c}{Target AMD Methods}                                                                                                                                                                                                \\ \cmidrule(l){3-7}
		                                                                                   &            &    Drebin      &   Drebin-DL   &    APIGraph      &  \multicolumn{2}{c}{MaMadroid}                                                        \\ \cmidrule(l){3-7}
		                                                                                   &
		                                                                                   &
		\begin{tabular}[c]{@{}c@{}}SVM\end{tabular}                            &
		\begin{tabular}[c]{@{}c@{}}MLP\end{tabular}                         &
		\begin{tabular}[c]{@{}c@{}}SVM\end{tabular}                           &
		\begin{tabular}[c]{@{}c@{}}RF\end{tabular}                            &
		\begin{tabular}[c]{@{}c@{}}3-NN\end{tabular}                                  \\ \midrule
		\multirow{4}{*}{AdvDroidZero}                                              & 10       & 45                      & 57                      & 43 & 92 & 74  \\
		                                                                                   & 20  & 60                      & 56                      & 56 & 88 & 84  \\
		                                                                                   & 30      & 71                      & 63                               & 65          & 95 & 88   \\
		                                                                                   & 40           & 74                      & 58                               & 71 & 98 & 85  \\
		                                                                                   \midrule
		\multirow{4}{*}{MAB}                                                            & 10       & 37                               & 38                               & 24          & 85          & 81           \\
		                                                                                   & 20  & 39                              & 55                               & 38          & 91          & 86           \\
		                                                                                   & 30     & 44                               & 54                              & 41          & 93         & 82          \\
		                                                                                   & 40        & 53                             & 54                              & 41          & 94         & 82               \\
		                                                                                   \midrule
		\multirow{4}{*}{RA}                                                               & 10       & 28                               & 36                              & 24          & 85         & 79             \\
		                                                                                   & 20  & 44                               & 53                              & 39          & 90         & 82                   \\
		                                                                                   & 30      & 53                               & 49                              & 44         & 94         & 96                \\
		                                                                                   & 40      & 54                               & 56                            & 40         & 90          & 93                \\
		                                                                                   \bottomrule
	\end{tabular}
 
	\label{tab:attackeffectActualNumber}
\end{table*}

\end{document}